\documentclass[structabstract]{aa}  
\usepackage{graphicx}
\usepackage{rotating}
\usepackage{longtable,lscape}
\usepackage{txfonts}
\begin{document}
   \title{The Monitor Project: Stellar rotation at 13~Myr}

   \subtitle{I. A photometric monitoring
    survey of the young open cluster h~Per.}

  \author{E. Moraux\inst{1} \and S. Artemenko\inst{2} \and
    J. Bouvier\inst{1}, J. Irwin\inst{3} \and M. Ibrahimov\inst{4}
    \and T. Magakian\inst{5} \and K. Grankin\inst{2} \and
    E. Nikogossian\inst{5} \and C. Cardoso\inst{6} \and
    S. Hodgkin\inst{7} \and S. Aigrain\inst{8} \and
    T.A. Movsessian\inst{5}}

   \institute{UJF-Grenoble 1 / CNRS-INSU, Institut de Planétologie et
     d'Astrophysique de Grenoble (IPAG) UMR 5274, Grenoble, F-38041,
     France
     \and Crimean Astrophysical Observatory, p/o Nauchny, 98409,
     Crimea, Ukraine
     \and Harvard-Smithsonian Center for Astrophysics, 60 Garden
     Street, Cambridge, MA 02138, USA
     \and Ulugh Bek Astronomical Institute of the Uzbek Academy of
     Sciences, Astronomicheskaya 33, 700052, Tashkent, Uzbekistan
     \and Byurakan Astrophysical Observatory, 0213, Aragatsotn reg.,
     Armenia
     \and Istituto Nazionale di Astrofisica, Osservatorio Astronomico
     di Torino, Strada Osservatrio 20, 10025 Pino Torinese, Italy
     \and Institute of Astronomy, University of Cambridge, Madingley
     Road, Cambridge, CB3 0HA, UK
     \and Astrophysics, Denys Wilkinson Building, University of
     Oxford, Oxford OX1 3RH, UK
            }

   \date{Accepted 20/06/2013}

 
  \abstract
   {}
   {We aim at constraining the angular momentum evolution of low mass
   stars by measuring their rotation rates when they begin to evolve
   freely towards the ZAMS, i.e. after the disk accretion
   phase has stopped.}
 {We conducted a multi-site photometric monitoring of the young open
   cluster h Persei that has an age of $\sim$13 Myr. The observations
   were done in the $I$-band using 4 different telescopes and the
   variability study is sensitive to periods from less than 0.2 day to
   20 days.}
 {Rotation periods are derived for 586 candidate cluster members over
   the mass range $0.4 \le M/M_{\odot} \le 1.4$. The rotation period
   distribution indicates a sligthly higher fraction of fast rotators
   for the lower mass objects, although the lower and upper envelopes
   of the rotation period distribution, located respectively at
   $\sim$0.2-0.3d and $\sim$ 10d, are remarkably flat over the whole
   mass range. We combine this period distribution with previous
   results obtained in younger and older clusters to model the angular
   momentum evolution of low mass stars during the PMS.}
 {The h Per cluster provides the first statistically robust estimate
   of the rotational period distribution of solar-type and lower mass
   stars at the end of the PMS accretion phase ($\ge$10 Myr). The
   results are consistent with models that assume significant
   core-envelope decoupling during the angular momentum evolution to
   the ZAMS.}

   \keywords{Stars: rotation -- Stars: low-mass -- open clusters and
     associations: individual: h Per}

\authorrunning{Moraux et al.}
\titlerunning{Stellar rotation at 13 Myr}
   \maketitle
%

\section{Introduction}

The angular momentum evolution of low mass stars is best characterized
by measuring the rotation rates of coeval populations at well-defined
evolutionary stages. Star forming regions and young open clusters
offer such an opportunity for the pre-main sequence (PMS) and main
sequence (MS) evolutionary phases, as they harbour rich stellar
populations of known age. Thus, in recent years, thousands of
rotational periods have been derived for low-mass stars and brown
dwarfs in these environments, covering an age range from 1~Myr to
1~Gyr. A compilation of these efforts was provided by Irwin \& Bouvier
(2009) and rotational distributions at additional age steps have been
further documented since then (e.g., Hartman et al. 2010; Littlefair
et al. 2010; Meibom et al. 2011; Affer et al. 2013). These results
yield prime observational constraints for the developement of stellar
angular momentum models (e.g., Gallet \& Bouvier 2013). Yet, a major
deficiency on the time sampling provided by these studies concerns the
end of the pre-main sequence evolution, between 5~Myr (Irwin et
al. 2008a) and 40~Myr (Irwinet al. 2008b), which remains poorly
documented (Messina et al. 2010, 2011). The goal of the study
presented in this paper is to remedy the lack of statistically robust
rotational distributions in this age range by measuring the rotational
periods of hundreds of low-mass members of the young open cluster
h~Per at an age of about 13~Myr (Mayne \& Naylor 2008).

H Per, also known as NGC 869, is the westernmost of the Perseus double
cluster with $\chi$ Per (or NGC884), separated by about 30'. Visible
to the naked eye and probably known since antiquity, this region was
first cataloged by Hipparchus as a patch of light in the Perseus
constellation in $\sim$130 B.C..  Its true nature, however, was only
discovered in 1780, well after the telescope invention, when sir
William Herschel recognized h \& $\chi$ Persei as two separate stellar
clusters. Since early 1900, it has been the target of extensive
photometric surveys using photographic plates (e.g. van Maanen 1911;
Oosterhoff 1937) and spectroscopic studies (e.g. Trumpler 1926;
Bidelman 1943; Schild 1965), leading to some controversy about the
distances and relative ages of h and $\chi$ Per. More recent studies
(e.g. Keller et al. 2001; Capilla \& Fabregat 2002; Slesnick et
al. 2002; Uribe et al. 2002; Bragg \& Kenyon 2005; Mayne et al. 2007;
Currie et al. 2010) are now converging to similar properties for both
clusters, with a distance modulus of $\sim11.8$, an extinction
$E(B-V)\sim0.54$ and an age of $\sim 13$ Myr (Mayne \& Naylor 2008). H
Per is about 30\% more populous than $\chi$ Per with at least
$\sim3000$ stars within its 10' center and has an estimated mass of
$\sim 4700 M_{\odot}$ (Currie et al. 2010).

The age of h~Per samples a critical phase for angular momentum
evolution models. By that age, the disk accretion process has ceased,
and the stars begin to freely evolve towards the zero-age main
sequence (ZAMS). Hence, deriving the rotational distribution of
low-mass stars in this cluster allows us to characterize the angular
momentum properties of stars at the end of the accretion phase, when
they have finally acquired their total mass and their angular momentum
evolution is not affected any more by the interaction with the
circumstellar disk. The distribution of rotation rates at this age
also provides the initial conditions for further evolution to the ZAMS
and onto the MS.

The paper is organized as follows. In the next section we describe the
observations and explain briefly how the light curves were produced
for selected possible candidate members. In Section 3, we present the
method we used to measure rotational periods and discuss possible
biases and contamination before giving our results. A general
discussion on the angular momentum PMS evolution follows in Section 4
and Section 5 summarizes our conclusions.


\section {Observations and light curve production of possible cluster members}

\begin{figure}
\centering
\includegraphics[width=0.5\textwidth]{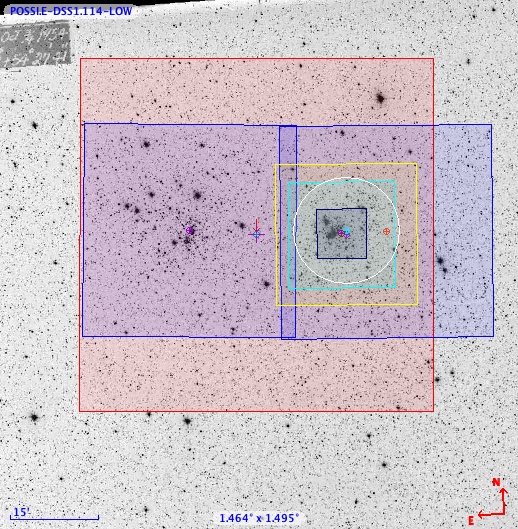}
\caption{Pointings of the various instruments used for this monitoring
  campaign: CFHT/MegaCam (red square), Maidanak (cyan square), CrAO,
  Byurakan (white circle). The field of view
  observed in the optical by Currie et al. (2010; blue rectangles) and in the
  near-infrared with CFHT/WIRCam by Cardoso et al. (in prep.; yellow square) are also
  shown. Their photometry is used to assess membership (see
  section~\ref{sec:sample}).}
\label{fov}
\end{figure}

\begin{figure}
\centering
\includegraphics[width=0.5\textwidth]{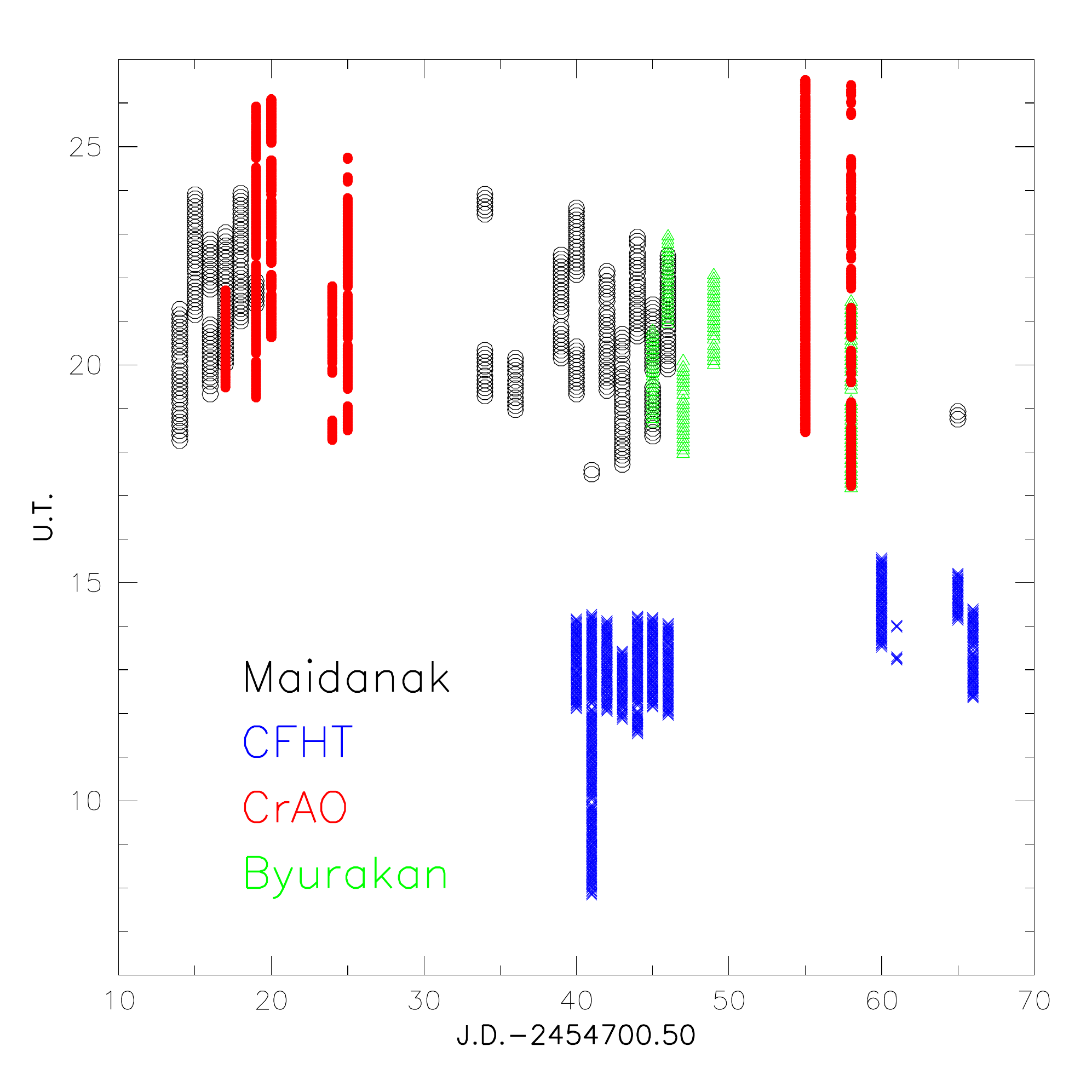}
\caption{Temporal sampling of photometric measurements obtained at
  Mt. Maidanak (black open circles), CFHT (blue crosses), CrAO (red dots),
and Byurakan (green triangles) observatories.}
\label{sampling}
\end{figure}

Within the framework of the Monitor project (Aigrain et al. 2007), we
conducted a multi-site photometric monitoring of h Persei during the
fall 2008 using 4 different telescopes: the 3.6m Canada-France-Hawaii
Telescope (CFHT), the 1.5m telescope in Maidanak (Uzbekistan), the
2.6m Shajn telescope (ZTSh) at the Crimean Astrophysical Observatory
(CrAO, Ukraine), and the 2.6m telescope of Byurakan observatory
(Armenia). The total amount of time spend to monitor the cluster was
$\sim110$ hrs, spread over 2 months from Sept. 5 to Oct. 27, 2008. The
observations were done in the $I$-band and the individual exposure
times were adapted to reach the equivalent of $i'_{CFHT}\simeq21$ with
a signal to noise larger or equal to 10 at each telescope. The details
of the observations performed at each telescope are described below.
Table~\ref{obs} provides a summary of all the observations,
Figure~\ref{fov} represents the various field of views and
Figure~\ref{sampling} shows the sampling rate of the light curves.

\begin{table*}[t]
  \caption{Journal of the observations}
 \begin{tabular}{llllllll}
    Telescope & pointing & FOV & Dates of & Number & Individual & Total number & Total \\
    &&& observations & of nights & exp. time & of exposures & hours \\
    \hline
    CFHT 3.6m & 02:20:42 +57:08:00 & $1\degr\times1\degr$ & Oct. 1-7 & 7 & 75s (+20s) & 698 (+55) & 23h \\
    &&& Oct. 21-27 & 3 &  &  & \\
    \hline
    Maidanak 1.5m & 02:18:56 +57:08:00 & $18'\times18'$ & Sep. 5-9 & 5 & 300s & 402 & 42h \\
    &&& Sep. 25 - Oct. 7 & 10 &  &  & \\
    &&& Oct. 25-26 & 2 & & & \\
    \hline
    CrAO 2.6m & 02:18:56 + 57:08:10 & $8.4'\times8.4'$ & Sep 8-16 & 5 & 120s & 992 & 33h \\
    &&& Oct. 16-17 & 2 & & & \\
    \hline
    Byurakan 2.6m& 02:18:50 +57:08:45 & 9' radius & Oct. 6-19 & 5 & 120s & 142 & 12h \\
    \hline
    Total &&& Sep. 5 - Oct. 27 & 27 & & 2234 & 110h \\
  \end{tabular}
 \label{obs}
\end{table*}

\subsection{Instrumentation and observing strategy} 
\label{sec:instrumentation}

At CFHT, the observations were performed in the $i'_{CFHT}$
filter with the 1deg$\times$1deg MegaCam CCD mosaic (Boulade et
al. 2003) over two runs (runID: 08BF06, P.I. E. Moraux and runID:
08BH17, P.I. B. Reipurth). Observing blocks of $\sim2$h were obtained
over 10 nights in Oct. 1-7 and Oct. 21, 26, 27, 2008, yielding a total
observing time of 23h. In total, 698 individual exposure times of 75s
as well as 55 short exposures of 20s were obtained. The observing
conditions were photometric and obtained during dark or grey time. The
seeing was between 0.6 and 1 arcsec. One of the longest observing
nights includes continuous measurements over more than 6h (0.25d),
which proved very useful to distinguish between long and short period
aliases in the period search analysis (cf. Section~\ref{sec:periods}).

At Maidanak, the observations were obtained in the $I_{Maid}$-band filter
using the SI 600 Series 4k$\times$4k CCD camera at the 1.5m AZT-22
telescope. The field of view of the camera is 18'$\times$18'. The
images are round and good up to 15 arcmin from the center but there is
a small coma-like distortion at the edge of the FOV. The individual
exposure time was 300s. A total of 42.3 observing hours (corresponding
to 402 individual frames) divided into blocks of $\sim$3 hours were
obtained over 3 different periods (Sep. 5-9, Sep. 25-Oct. 7 and
Oct. 25-26). Only the first two periods benefited from good weather
conditions and the data obtained during the 3rd period are not used in
our analysis.

At CrAO, the observations were carried out in the $I_{CrAO}$ photometric
band, using a FLI PL-1001E CCD camera at the prime focus of the 2.6-m
Shajn telescope (ZTSh). The pixel size of the 1024 x 1024 size CCD
matrix KAF-1001E is 0.5 arcsec and the field of view is
8.4'$\times$8.4'. The cluster was imaged during 7 nights on Sept.
8, 10, 11, 15, 16 and Oct. 16, 19, 2008. The exposure time for each
frame was 120 seconds and the total time of observations was about 33
hours corresponding to $\sim$5 hours per night,  yielding a total of
992 images. As weather conditions on September 15-16 were bad, we
have excluded these two nights from the subsequent analysis.

At Byurakan, the observations were done at the prime focus of the 2.6m
telescope with the ByuFOSC-2 camera, equipped with 2058 x 2063 pix
Loral CCD matrix. This equipement provides a scale of 0.5 arcsec/pix
and a full view of 17'$\times$17'. However, due to the absence of
correcting optics, the non-distorted FOV is reduced to about
9'$\times$9'. Images were taken through the $I_{Byur}$-band filter with an
exposure time of 120 sec. The seeing was about 2 arcsec during the
observations performed over 5 nights on Oct. 6, 7, 8,
10, and 19, and the observing blocks were 2 hours long except on Oct. 19
where it was 4 hours long, yielding a total of 142 images.

\subsection{Light curve production}
\label{sec:light-curve}

The data processing and light curve production were performed for each
dataset separately following the procedure described in Irwin et
al. (2007a). Briefly, a master frame was created by stacking the images
taken in the best seeing conditions before using the source detection
software. The resulting master catalog containing the positions of the
detected objects was then used to perform aperture photometry on all
the individual images. For each frame, the difference between each
source magnitude and its median (calculated across all frames) was
measured to compute the frame offset and correct for seeing
variation. A 2-D fit of the residual map was then subsequently removed
to account for varying differential atmospheric extinction across each
frame. Light curves were extracted for $\sim$219 000, $\sim$8000,
$\sim$2000 and $\sim$3000 objects from the CFHT, Maidanak, CrAO and
Byurakan images.

\begin{figure}
  \centering
  \includegraphics[angle=270,width=0.45\textwidth]{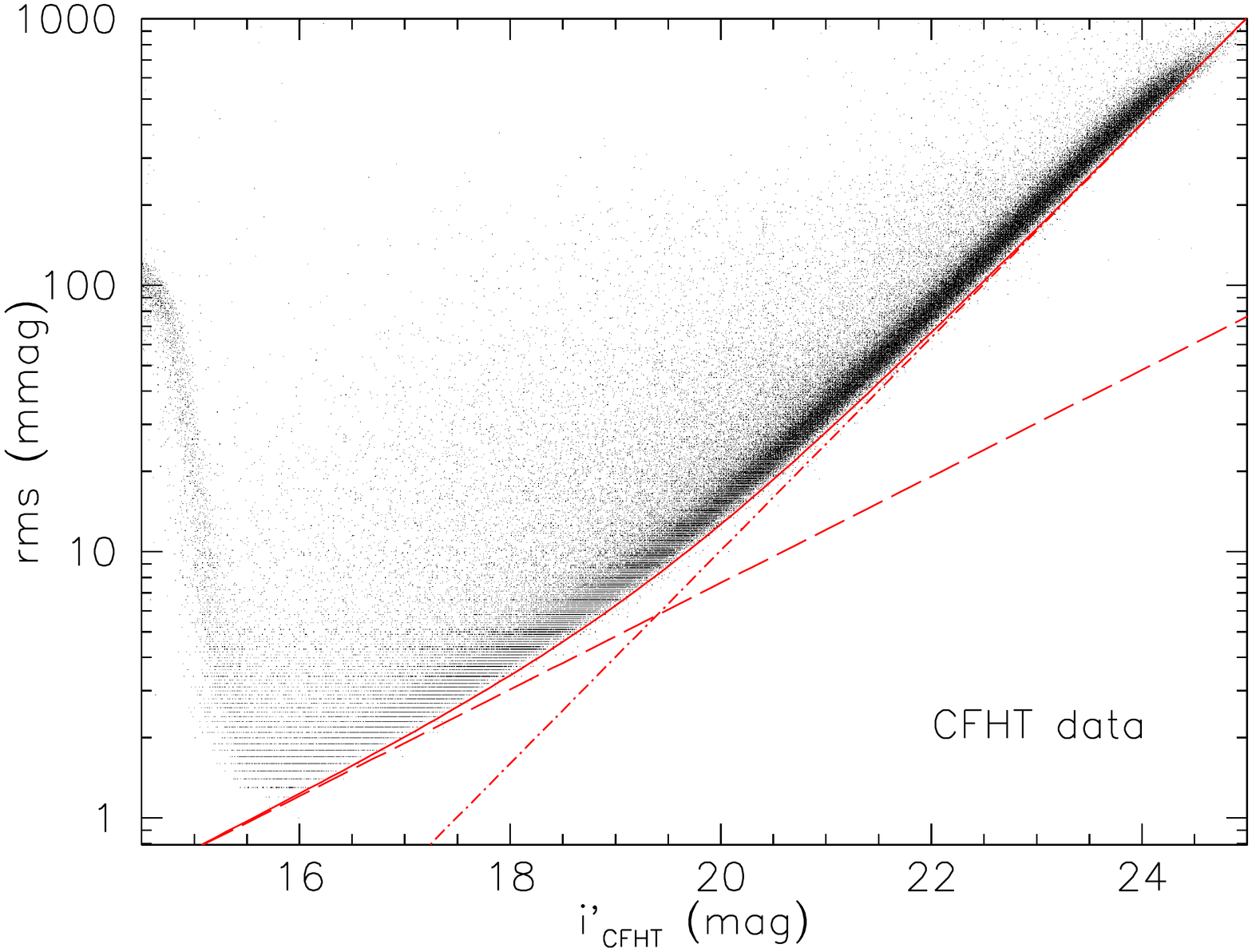}\\
  \includegraphics[angle=270,width=0.45\textwidth]{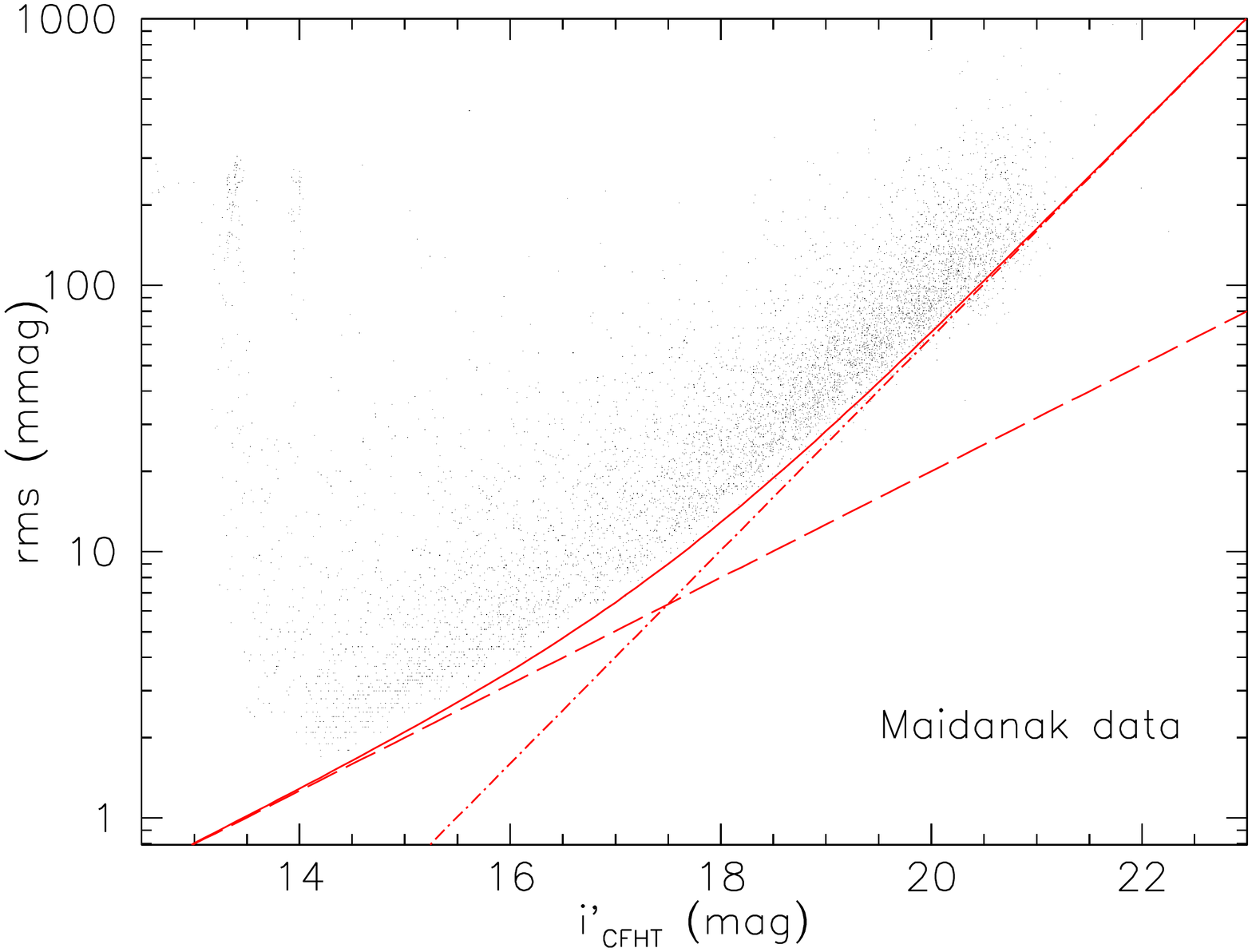}
  \caption{Plot of RMS scatter per data point measured over the CFHT
    and Maidanak dataset as a function of $i'_{CFHT} $ magnitude. The
    diagonal dashed lines show the expected RMS from Poisson noise in
    the object, the diagonal dot-dashed lines show the RMS from sky
    noise in the photometric aperture. The solid lines show the
    overall predicted RMS, combining these contributions.}
  \label{rms}
\end{figure}

For the CFHT dataset, the achieved photometric precision for each data
point is very good: it is better than 2 mmag for the brightest objects
($i'_{CFHT} \le 16$), with a scatter $<1$ per cent up to
$i'_{CFHT}\simeq19.5$ (see Fig.~\ref{rms}). The detection limit on
a single frame corresponding to a signal-to-noise ratio of 5 is
reached at $i'_{CFHT}\simeq23.5$. The photometric precision for the
Maidanak data is also very good, although not as deep.  For the CrAO
and Byurakan photometry however, the rms scatter never gets better
than 5 mmag even for the brightest objects.

For each possible cluster member (see next section for a description
of the selection),
a quality check of each data point was performed before combining the
light curves of the 4 different sites. If its magnitude was brighter
than $i'_{CFHT}<15.6$ ($i'_{CFHT}<13.8$) for the long (short)
exposures of the CFHT images or $i'_{CFHT}<14$ for the other data, we
considered it as saturated (see Fig.~\ref{rms}). If the aperture contained bad pixels, if
the frame seeing was extremely bad or if the frame offset was very
large (typically if it falls in the worst 5\% of the distribution in
both cases), the data point was qualified as bad. We also marked all
the objects located more than 550 pixels (or 4.6 arcmin) away from the
center of Byurakan images due to strong distortion. And finally, we
used a sigma clipping to identify all the data points for which the
magnitude is more than 3 sigma away from the average magnitude of the
10 closest data points. In all cases, a flag was added to the data
point and we kept only those with a flag 0 for our analysis.

For each object, we then computed the median magnitude of each light
curve using only the good data points. We normalised the obtained
value to the CFHT median taken as a reference by applying an offset,
and we combined the light curves. No color term was applied. For the
period search described in section 4, we used only the data from CFHT
and Maidanak however since the photometric precision is best for these
two datasets. The data from CrAO and Byurakan observatory were used
{\it a posteriori} to check the goodness of the period found.

\subsection {Possible cluster member selection}
\label{sec:sample}

We first built a master catalog by cross correlating all the sources
detected in any of the monitoring images described above, i.e. from
either CFHT, Maidanak, CrAO or Byurakan observatory. Since the field
of view and the image quality was better for the CFHT and Maidanak
datasets, we decided to focus on the objects that were detected at
least in one of these two. The cross-correlation was done by
coordinates with TOPCAT (Taylor 2005) using a matching radius of 2
arcsec. This leads to a catalog with 219,642 sources with $i'_{CFHT}
\simeq 13-24$.

We then cross-correlated our master catalog with data from the
literature (Slesnick et al. 2002, Uribe et al. 2002, Mayne et
al. 2007, Currie et al. 2007a, 2007b, 2007c, 2008 and 2010) in order
to get as many photometric and/or spectroscopic information for each
source. In particular, we used the $V$ and $I_C$ magnitude from Currie
et al. (2010) as the observed field of view is very similar to our
MegaCam pointing (see Fig.~\ref{fov}). We also used YJHK CFHT/WIRCam
data (Cardoso et al., in prep.)  as well as Chandra X-ray data
(C. Argiroffi, priv. com.). Again the matching radius was set to 2
arcsec, except for the X-ray data for which we used 2.5 arcsec.

\begin{figure*}
\centering
\includegraphics[angle=270,width=\textwidth]{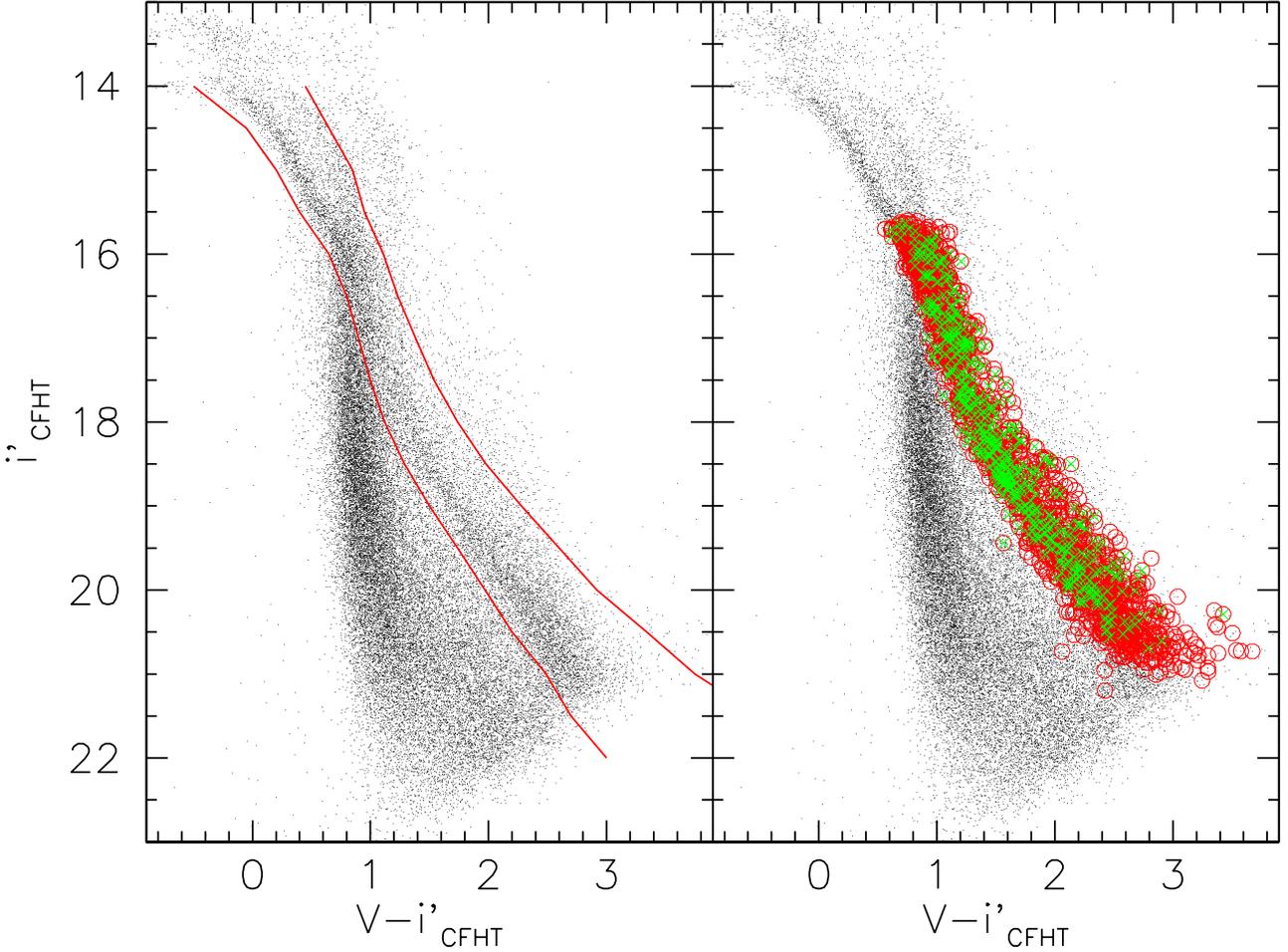}
\caption{$i'_{CFHT},V-i'_{CFHT}$ colour magnitude diagram. {\it Left:} All the
  objects located between the two solid lines have been selected as
  possible cluster members. {\it Right:} The red open circles show the
  objects that have been analysed while the green crosses show the
  periodic objects (a colour version is available on the journal electronic
  version).}
\label{cmd_VI}
\end{figure*}

Using all the available photometric data we built colour-magnitude
diagrams. Since the cluster sequence is clearly visible in all the
CMDs, we defined an empirical sequence to assess
membership. As an example, we can see in Fig.~\ref{cmd_VI} the
selection that we made in $i'_{CFHT}, V-i'_{CFHT}$. The two solid lines represent
the custer sequence shifted in colours by $\pm0.1$ mag plus the amount
of the photometric error, and by -0.75 mag in $i'_{CFHT}$ for the upper
envelope to take into account binaries. All the objects located
between the two lines were selected as candidate members. We performed
the same selection on all the available CMDs. An object was eventually
considered as a possible cluster member if its optical $V-i'_{CFHT}$ colour
and/or near-infrared WIRCam colours ($J-H$, $J-K$ and $H-K$) were
consistent with membership. Alternatively, if these colours were not
available (no detection) but the object was detected in H$\alpha$
(Currie et al. 2007c), or in X-ray (Currie et al. 2009, and
C. Argiroffi priv. com.), or was classified as possible member by
Currie et al. 2010 (membership flag $\ge 1$) or Uribe et al. 2002
based on proper motion, we kept it as a cluster candidate. We ended up
with a rather conservative final catalog of 19,224 sources selected as
possible members over the whole field of view. 45,710 objects were
rejected based on their photometric colours and 154,708 objects were
classified as ``unknown'' since we did not have any membership
information (objects detected only in one filter, usually with MegaCam
at CFHT since its FOV is larger than any other h Per survey that
has been performed so far).

\begin{figure}
\centering
\includegraphics[width=0.5\textwidth]{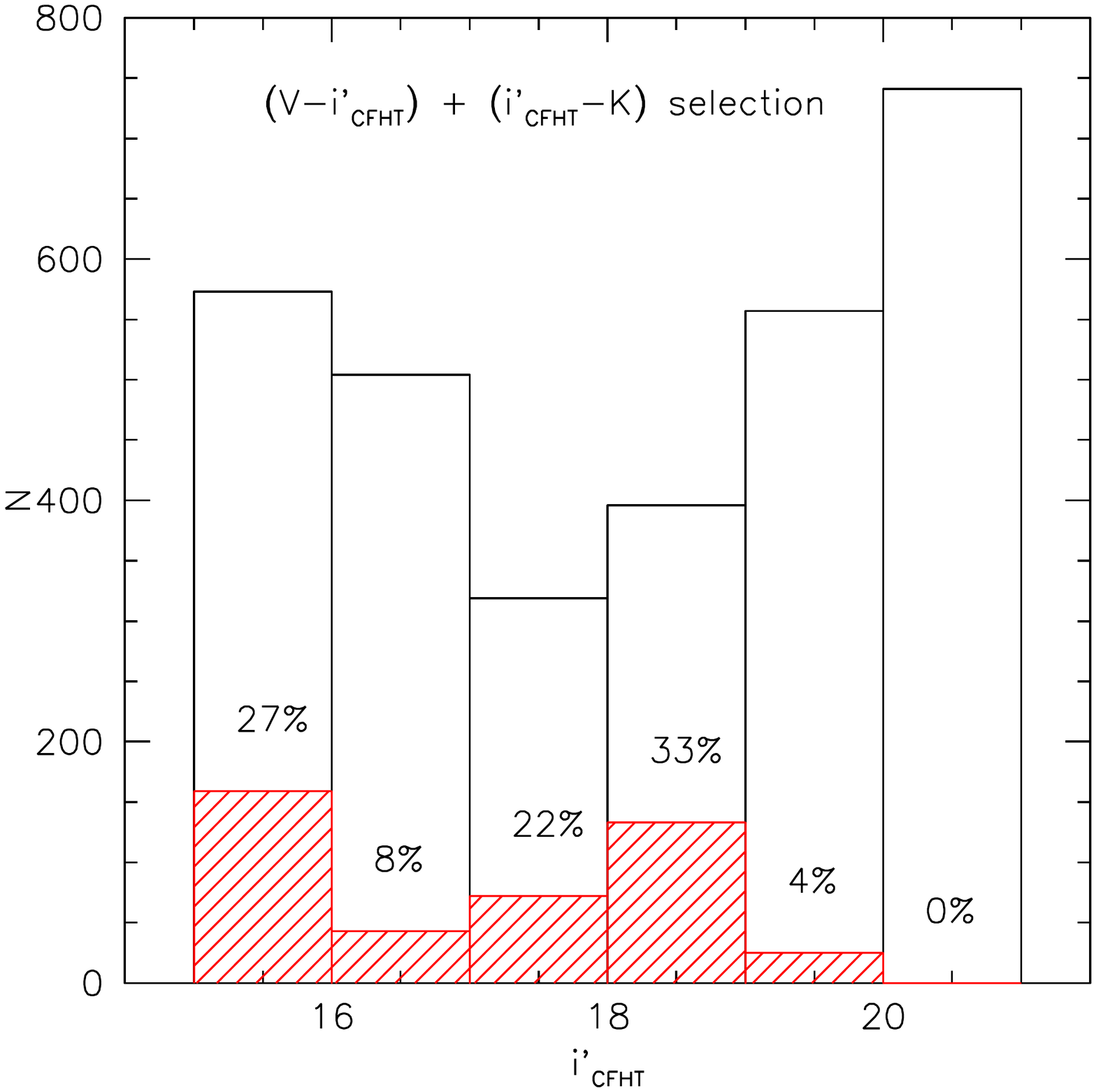}
\caption{Photometric contamination level per magnitude bin using the
  Besancon galactic model after selection in both $V-i'_{CFHT}$ and $i'_{CFHT}-K$
  colours. The plain histogram corresponds to the number of objects
  selected as possible members from our data (and located in the WIRCAM
  FOV), while the dashed one corresponds to the selection performed on
  the synthetic catalogue obtained from the Besancon model for the
  same WIRCAM pointing.}
\label{contamination}
\end{figure}

In order to estimate the level of contamination by field objects in
our photometric selection, we used the Besancon galactic model (Robin
et al. 2003). We performed the same analysis on our data and on the
synthetic catalog to which photometric noise has been added. In both
cases the considered FOV was restricted to the WIRCam pointing and we
used only two CMDs ($i'_{CFHT}, V-i'_{CFHT}$ and $i'_{CFHT},
i'_{CFHT}-K$) for the selection. This will only provide us an upper
limit of the contamination as we actually used more colours for the
final photometric selection but this is sufficient for our purpose,
especially as these two colours are the most constraining for
membership selection. Overall we found a contamination level of 14\%
after selection on both $V-i'_{CFHT}$ and $i'_{CFHT}-K$ colours (26\%
if only $V-i'_{CFHT}$ is used, 34\% if only $i'_{CFHT}-K$) in the
magnitude range $i'_{CFHT}=15-21$. The contamination per magnitude bin
is given in Fig.~\ref{contamination}. It is interesting to note that
the selection drops to zero for the faintest objects. This is due to
the fact that the cluster members are much redder than the galactic
population in the chosen CMDs for $i'_{CFHT}=20-21$ (see
Fig.~\ref{cmd_VI}). There is therefore no contamination by field stars
in this magnitude range (but by galaxies). For brighter magnitudes,
however, the contamination gets larger especially when the cluster
sequence joins the dwarf locus.

For the rest of the paper, we restricted our study to the possible
members with $i'_{CFHT}>15.6$ (16,381 objects) to avoid saturation on
the CFHT long exposures (the time sampling on the short exposures
being too low).  Following the work from Cardoso et al. (in prep.), we
converted the $i'_{CFHT}$ magnitude of each candidate member into mass
using the mass-magnitude relationship from Siess model at 13 Myr
(Siess et al. 2000).


\section{Period measurement analysis}

\subsection{Period search}
\label{sec:periods}

In order to take the best benefit of our sampling, we performed the
period search on objects detected in both CFHT and Maidanak images,
i.e. with light curves containing more than 800 measurements (and up
to 986). This allows us to be sensitive to both short and long
periods and to avoid confusion between harmonics as much as
possible. The CFHT and Maidanak measurements are spread over 51
nights, from JD 2454716 to JD 2454767, with two large observing gaps
from JD 2454720 to JD 2454735 and from JD 2454748 to JD 2454761. The
sampling rate is graphically shown in
Figure~\ref{sampling}. Photometry obtained at other sites is of
lower quality and was only used {\it a posteriori} to confirm the measured
periods. Of the 16,381 h Per photometric candidate members located within
the CFHT FOV with $i'_{CFHT}>15.6$, only 2287 were detected in both
CFHT and Maidanak data. We thus restricted our period analysis to
these sources. 

Periodic signals were searched in the light curves using 3 methods :
Lomb-Scargle periodogram (Scargle 1982; Horne \& Baliunas 1986), CLEAN
Discrete Fourier Transform (CLN DFT, Roberts et al. 1987) and
String-Length minimization (Dworetsky 1983). In all cases, we probed a
range of frequency from 0.05 to 5.0~d$^{-1}$ (i.e., periods
from 0.2 to 20 days). The lower frequency limit is about twice the
frequency resolution, $\delta$f$\simeq$1/T where T is the total time span of
the observations. The higher frequency limit was set somewhat arbitrarily to
reduce computation time, though we are sensitive to much shorter
periods ($f_{max}\simeq \delta t^{-1}$$\simeq$20~d$^{-1}$,
see below). The frequency range was explored by step of
0.001~d$^{-1}$. At every sampled frequency, the power of the
periodogram and of the CLEAN DFT, as well as the string-length were
computed. For each light curve, we thus derive 3 possible periods
corresponding to the highest peaks in the periodogram and CLN DFT, and
to the shortest string-length in the phase diagram.

\begin{figure}
\centering
\includegraphics[width=0.5\textwidth]{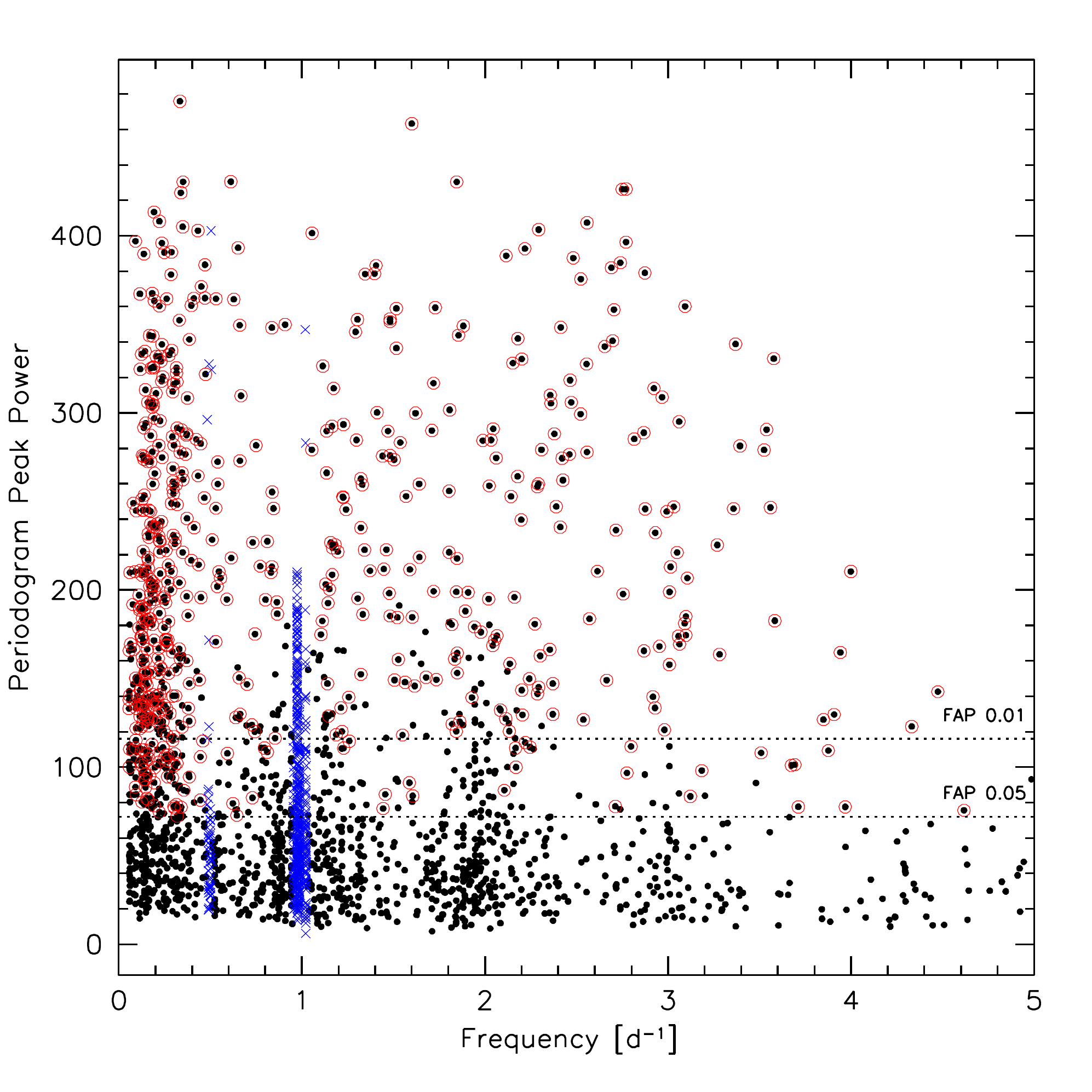}
\caption{The distribution of the power of the highest periodogram peak
  is shown as a function of frequency for 2287 h~Per photometric
  candidate members analyzed for periodicity. Spurious periods
  occurring at frequencies around 0.5 and 1.0d$^{-1}$ (shown as blue
  crosses) were discarded. The FAP levels computed from a control sample
  of non-periodic field stars are indicated. The 586 h~Per candidate
  members that were eventually considered as periodic are encircled
  (red).}
\label{perpeak}
\end{figure}

The power of the highest peak recorded in the periodogram of each of
the 2287 light curve is plotted as a function of frequency in
Figure~\ref{perpeak}. A clear accumulation of peaks occurs at
frequencies around 0.5 and 1d$^{-1}$. These peaks correspond to
spurious periods resulting from the nightly sampling rate. We
therefore removed all periods lying in the frequency ranges 0.48-0.51
and 0.95-1.03d$^{-1}$. This left us with 1761 possibly periodic h~Per
candidate members.  As the investigated light curves have nearly the
same number of measurements and a similar temporal sampling, the
highest peak power also corresponds to the lowest False Alarm
Probability (FAP, cf. Scargle 1982).

We derived FAP levels from a control sample (discussed in more details
in Section~\ref{sec:control}) of 1307 field stars, i.e., photometric
{\it non} members spanning a similar magnitude and color ranges in the
($V, V-i'_{CFHT}$) color-magnitude as h Per photometric candidate
members. The light curves of these 1307 field stars also have a
similar temporal sampling as those of the h~Per candidate members
under analysis. The cumulative distribution of the power of the
highest peak in the periodograms of the field stars provides an
estimate of the FAP, i.e., the probability that a given peak power
merely results from statistical noise. After removal of spurious peaks
at frequencies around 0.5 and 1.0~d$^{-1}$ (see above), we are left
with 890 field stars, from which we derive a FAP level of 0.01 for a
peak power of 116, and FAP of 0.05 for a peak power of 72. These FAPs
are illustrated in Fig.~\ref{perpeak}.  As discussed by Littlefair et
al. (2010), this approach is more accurate than analytical estimates
of the FAP and/or FAP derived from Monte Carlo simulations.

Of the 1761 h~Per candidates under analysis, we retained 872 with a
FAP level less than 0.05 (604 with a FAP $<0.01$) as being possibly
periodic\footnote{Note that using a much lower FAP level, as
  sometimes done in other studies, may introduce a bias against long
  periods and small amplitudes.}. Their light curves were phase-folded
with period estimates derived from the periodogram, CLN DFT and
string-length methods. In most cases, the periodogram and CLN DFT
return similar results. Whenever the 3 methods yielded different
estimates, we visually relied on the phase diagrams to decide on the
best period, if any. CLN DFT proved very reliable to detect the
periods of sinusoidal-like light curves, while string-length was found
sometimes, though rarely, superior for more complex signals (e.g
double-spotted light curves, cf. HPer-195, -243, -431).  The FAP
analysis is not fully reliable on its own (cf. Littlefair et al. 2010)
and must be complemented by eye-inspection of the phased light curves
to remove interlopers. A significant fraction of the selected possibly
periodic candidate members did not show any convincing phased light
curves and we rejected 283 candidates as being non periodic. Among
them, 90 had a FAP less than 0.01 while the others (193) had a FAP
between 0.01 and up to 0.05. We considered that their phased light
curves did not show periodic modulation by surface spots for one of
the following reasons: unconvincing periods close to 0.5d or 1d (about
50\% of the rejected light curves have periods in the range 0.4-0.6d
or 0.8-1.2d); presence of superimposed low and high frequency
modulations, presumably frequency aliases, which do not allow us to
reliably assess the true period; discontinuous photometric jumps in
the phased light curves and/or systematic offsets between Maidanak and
CFHT data; significant discrepancies between CFHT and Maidanak phased
light curves; single flares or dips dominating the photometric
amplitude.

In order to test the effect of the chosen FAP level, we defined two
samples based on two different FAP cuts at 0.01 and 0.05, containing
respectively 514 and 586 periodic objects after eye examination.  We
then compared their period distributions for the 4 mass bins discussed
in section~\ref{sec:distribution}. The KS probability that both
distributions are drawn from the same population is larger than
99.99\% for the three higher mass bins, and is equal to 82.36\% for
the lower mass bin (04-0.6 M$_\odot$), indicating that there is no
significative difference between the two samples. Thus, choosing one
or the other cut in FAP level would not affect the results of the
paper, and we decided to use the less selective threshold (FAP$<0.05$)
allowing to select all the clear periodic variables -- in the line
with our previous Monitor studies from Irwin et al. (2007b, 2008a,
2008b, 2009).

We finally ended with a sample of 586 periodic h~Per candidate
members, with a spectral type spanning a range from $\sim$ F0 to
M6. It is interesting to note that we measured a rotation period for
about 50 F stars, which is of the same order than the number of such
periodic stars found by Saesen et al. (2010) who investigated the
variability of bright stars in $\chi$ Per in a similar
FOV. Table~\ref{phot} (available online) lists the periods and
photometric amplitudes of all the periodic objects. The rms error on
each period can be computed as $\delta P = \delta \nu \times P^2$,
where $\delta \nu$ = 0.0145d$^{-1}$ is the average sigma of the
gaussian fit to the highest peak in the CLN DFT. The amplitude of
variability was derived from a sinusoidal fit to the light curves
using FAMOUS\footnote{based on the frequency mapping FAMOUS,
  F. Mignard, OCA/CNRS, ftp://ftp.obs-nice.fr/pub/mignard/Famous}.
The phased light curves of the 586 periodic variables are shown in the
appendix, Fig.[A1].  In some cases, especially for objects with short
periods, the phased light curve provides clear evidence for phase
and/or amplitude variations over the time span of the observations,
indicative of spot evolution and/or surface differential rotation over
a timescale of weeks (see e.g. HPer-208, $i'_{CFHT}=15.72$, as an
extreme case).

\onllongtab{2}{
\begin{landscape}
\begin{longtable}{lcccccccccccccccc}
  \caption{\label{phot} List of periodic objects ordered by
    periodogram peak power. The binary flag is set to 2 if the object
    has been identified as a binary in the $i'_{CFHT}, i'_{CFHT}-K$
    CMD and to 1 otherwise. The $V$ and $I_C$ magnitudes as well as
    the membership flag (column before last) are from Currie et
    al. (2010). The near infrared photometry has been obtained with
    WIRCam at CFHT and is from Cardoso et al. (in prep.). The last
    column indicates whether the object has been detected in H$\alpha$
    (Currie et al. 2007b) and/or in X-rays (Currie et al. 2009;
    C. Argiroffi priv. com)} \\
\hline
HPer & RA  & DEC  &  $i'_{CFHT}$ & Period  &   Power  &   FAP  &  amplitude &  mass &  binary & V     & Ic   &   J   &    H   &   K  &   Memb. & Other  \\ 
-ID   &  (deg)     &  (deg)    &  (mag)       & (d)     &           &        &  (mag)     &  $M_{\odot}$ & flag   & (mag) & (mag) & (mag)&   (mag)&  (mag) & flag &  \\
\hline          
\endfirsthead
\caption{Continued.} \\
\hline
HPer & RA  & DEC  &  $i'_{CFHT}$ & Period  &   Power  &   FAP  &  amplitude &  mass &  binary & V     & Ic &   J   &    H   &   K  &   Memb. & Other  \\ 
-ID   &  (deg)  & (deg)  &  (mag)      & (d)     &           &        &  (mag)     &  $M_{\odot}$ & flag   & (mag) & (mag) &  (mag)&   (mag)&  (mag) &  flag &  \\
\hline
\endhead
\hline
\endfoot
\hline
\endlastfoot
  1      & 02:19:20.17   & +57:11:51.4   & 17.35     & 2.99     & 476.0   & $<$0.001 & 0.064   & 1.21   & 1      & --       & --       & 15.72   & 15.09   & 14.90   & --     & X           \\ 
  2      & 02:19:38.13   & +57:12:32.3   & 17.11     & 0.625     & 463.3   & $<$0.001 & 0.046   & 1.25   & 1      & 18.32   & 16.63   & 15.48   & 14.83   & 14.63   & 3      & X           \\ 
  3      & 02:18:11.37   & +57:14:40.8   & 18.21     & 1.64     & 430.6   & $<$0.001 & 0.061   & 1.03   & 1      & --       & --       & 16.41   & 15.67   & 15.46   & --     & --          \\ 
  4      & 02:18:17.31   & +57:07:35.3   & 17.02     & 2.85     & 430.5   & $<$0.001 & 0.037   & 1.27   & 1      & 18.10   & 16.63   & 15.48   & 14.85   & 14.70   & 3      & --          \\ 
  5      & 02:18:31.14   & +57:03:53.3   & 16.43     & 0.541     & 430.4   & $<$0.001 & 0.029   & 1.34   & 2      & 17.57   & 16.10   & 14.94   & 14.33   & 14.17   & 3      & X           \\ 
  6      & 02:18:19.07   & +57:02:28.5   & 16.85     & 0.361     & 426.4   & $<$0.001 & 0.055   & 1.29   & 2      & 18.16   & 16.40   & 15.10   & 14.40   & 14.20   & 3      & X           \\ 
  7      & 02:19:23.99   & +57:02:24.8   & 16.71     & 0.364     & 426.2   & $<$0.001 & 0.022   & 1.31   & 2      & 17.88   & 16.25   & 15.10   & 14.45   & 14.27   & 3      & --          \\ 
  8      & 02:18:54.21   & +57:00:30.6   & 18.14     & 2.97     & 424.3   & $<$0.001 & 0.049   & 1.03   & 1      & 19.70   & 17.75   & 16.10   & 15.35   & 15.15   & 3      & --          \\
\end{longtable}
\end{landscape}
}

\begin{figure}
\centering
\includegraphics[width=0.4\textwidth]{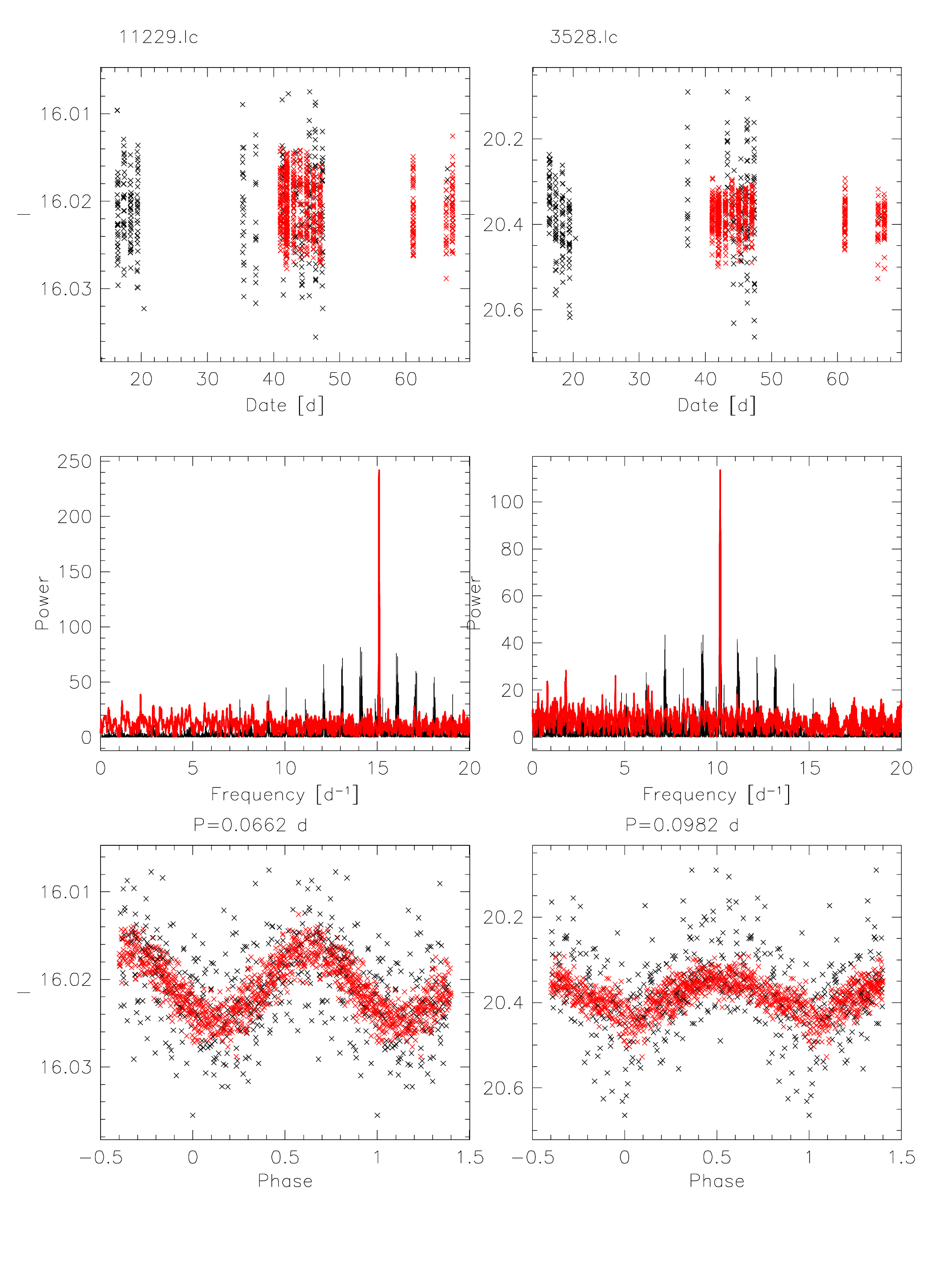}
\caption{The light curves of the 2 h~Per candidate members with the
  shortest photometric periods are shown in the top panels (red:
  CFHT, black: Maidanak), their frequency spectrum in the middle
  panels (black: periodogram, red: CLEAN DFT), and their phased
  light curves in the bottom panels (red: CFHT, black:
  Maidanak). {\it Left :} HPer-215 with a period of 0.06626~d; {\it
    Right :} HPer-513 with a period of 0.09827~d. }
\label{shortper}
\end{figure}

Note that some periods reported here are shorter than 0.2d, the lower
limit of our period search range. This is because visual inspection of
some phase diagrams clearly indicated that we had detected a lower
frequency harmonic of the true period.  In such a case, the period
search analysis was repeated with a lower limit of 0.05d. This yielded
5 objects with clear photometric periods shorter than 0.2d, with the 2
shortest ones, amounting to only 0.0662~d (1.59~h) and 0.0982~d
(2.36~h), illustrated in Figure~\ref{shortper}. The spectral type
( F1$\pm2.5$ according to Currie et al. 2010), mass
($\sim$1.4~M$_\odot$), and low photometric amplitude ($\sim$ 0.008
mag) of HPer-215 suggests $\delta$-Scuti type pulsations rather than
rotational modulation (Saesen et al. 2010; Chang et
  al. 2013). However, the red colors ($V-I_C\sim$4.5; Mayne et
al. 2007) of HPer-513, indicative of a spectral type $\sim$M6, put
this object far from expected pulsational instability strips at the
age of the cluster (Rodr\'{\i}guez-L\'opez et al. 2012; Baran et
al. 2011; Palla \& Baraffe 2005). The shape of the phased light-curve
is somewhat suggestive of a (P$_{orb}$=0.19654d) contact binary (Nefs
et al. 2012).

\medskip

The period analysis above was performed on the combined Maidanak and
CFHT datasets. The CrAO and Byurakan datasets, of lower quality and
with a sparser, though complementary, temporal sampling were 
used to visually confirm the results. In most cases, whenever the S/N
ratio of the latter datasets was high enough, the additional
measurements adjusted consistently, albeit with a much increased
scatter, onto the phased light curve, thus providing further support
to the reported periods.

\subsection{Period detection, completeness and reliability}
\label{sec:completeness}

\begin{figure*}
\centering
\includegraphics[width=0.9\textwidth]{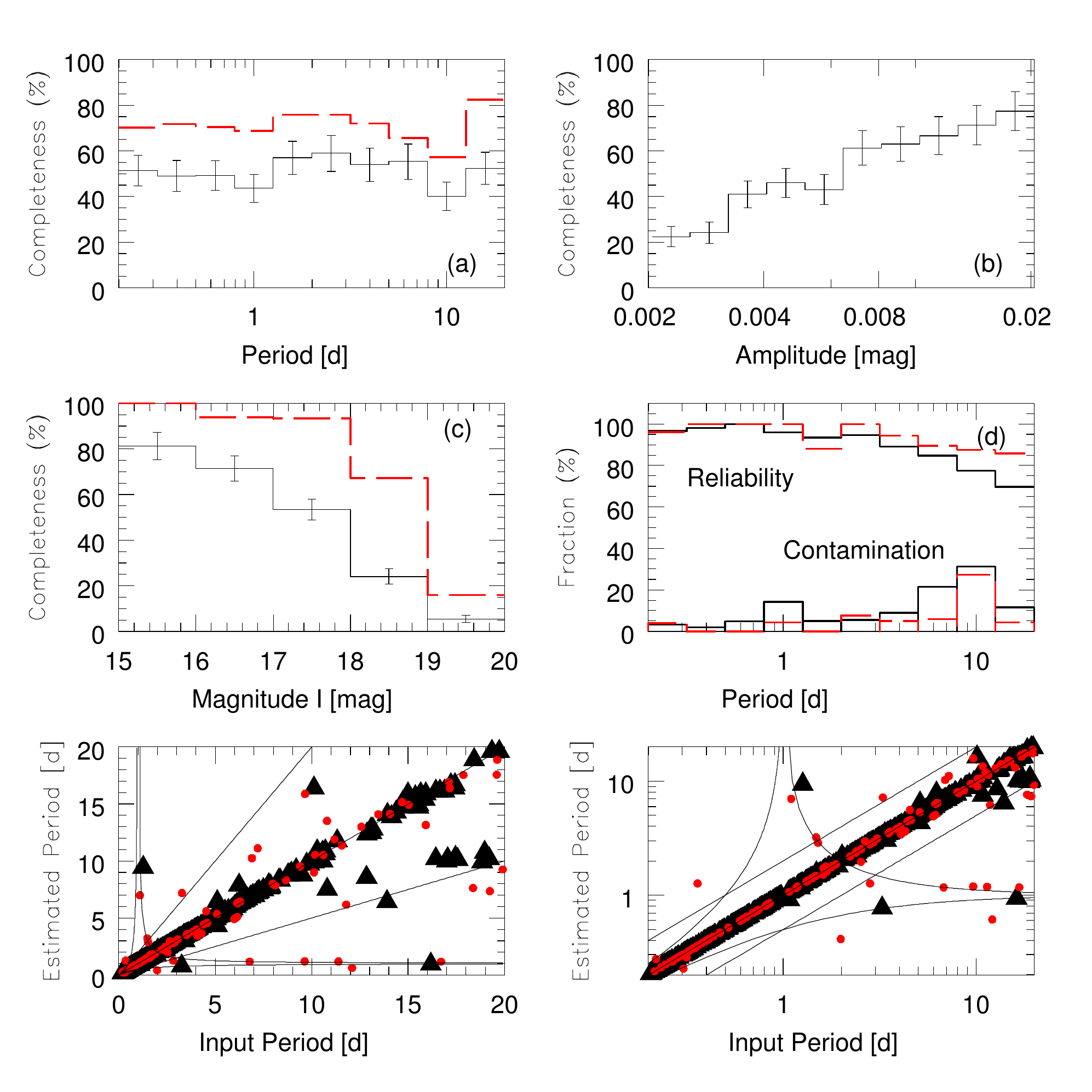}
\caption{Period detection completeness and reliability for sinusoidal
  signals with amplitudes from 0.002 to 0.02~mag (black histograms)
  and from 0.01 to 0.02~mag (red dashed histograms). The period
  detection completeness is shown as a function of period in panel
  (a), photometric amplitude in panel (b), and stellar magnitude
  in panel (c). The reliability of the recovered periods is shown in
  panel (d). Reliable periods are defined as those within 10\% of the
  input periods (shown as a function of input period). Contaminating
  periods are those that differ from the input period by more than
  10\% (shown as a function of detected period). The lower left and
  right panels show the estimated periods as a function of the input
  periods on a linear and log scales, respectively. Periods recovered
  with a FAP$\leq$0.01 are shown by large (black) triangles, while
  those with 0.01$<$FAP$\leq$0.05 are shown by small (red)
  circles. The straight lines show the locus of equal periods and factor of 2
  harmonics and the curves show the location of 1d period aliases. }
\label{completeness}
\end{figure*}
 
We describe here Monte Carlo simulations that we ran on synthetic
light curves to estimate the robustness and completeness of the period
search analysis presented in the previous section. In order to build
synthetic light curves, we first selected photometric cluster
non-members in the CMD, i.e., objects running below the lower envelope
of the cluster sequence, over the range $i'_{CFHT}\sim$15-21~mag. We
further selected a subset of these non-members, whose photometric
light curves had an rms dispersion lower than the median photometric
error at a given brightness. Of this subset, we retained only those
with light curves having at least 800 measurements from CFHT and
Maidanak, thus yielding a sampling rate similar to those of the h~Per
members analyzed above. This procedure resulted in a sample of 1065
light curves for non- (or low-) variable field stars.

To these light curves we then added a sinusoidal signal, with a
log-uniform distribution of periods in the range 0.2-20~d and of
amplitudes from 0.002 to 0.02~mag, and random phases. The
resulting light curves were subjected to exactly the same period
search analysis as described above for the candidate members. After
removing spurious periodogram peaks at frequencies around 0.5 and
1.0~d$^{-1}$, thus leaving 941 synthetic light curves, we retained 592
objects with a FAP less than 0.05 and vizualized their phased light
curves. We rejected 50 additional objects based on the large scatter
in their phased light curves. We were thus left with 542 periodic
object from this sub-sample, i.e., a global detection rate of 542/1065
= 51\%.

Comparing the detected periods to the input periods thus allows us to
estimate the period detection rate as a function of period, amplitude
and brightness, and to assess the sensitivity of the method to aliases
and harmonics. Note that the production of synthetic periodic light
curves has been done by one of the authors, and the search for periods
has been performed by another two who ignored the input parameters so
as not to bias the results.The results are shown in
Figure~\ref{completeness}. The completeness plot, i.e., the number of
detected periods divided by the number of input periods in a given
period bin, indicates that the period detection rate is relatively
uniform over the 0.2-20d range (Fig.~\ref{completeness}, panel
a). Figure~\ref{completeness} (panel b) also indicates that the period
detection rate is quite sensitive to the input amplitude, increasing
from about 20\% for amplitudes of 0.002~mag to nearly 80\% for
amplitudes of 0.02~mag. A sinusoidal fit to the light curves with
detected periods recovers the input amplitudes to within 1.1~mmag rms.
The period detection rate is equally sensitive to the stellar
magnitude (Fig.~\ref{completeness}, panel c), decreasing from
$\sim$80\% at $i'_{CFHT}\simeq$15.5 to a mere 5\% at
$i'_{CFHT}\simeq$19.5, where the photometric noise becomes comparable
to the injected amplitudes (see Fig.~\ref{rms}). We should stress,
however, that for the large surface spots expected for low mass stars
at the age of h~Per, the photometric amplitudes are likely to be
larger than $\sim$0.01~mag, thus alleviating the issue
of photometric noise for faint members.

It is quite noticeable that a large fraction (490/542=90\%) of the
recovered periods agree with the input ones to better than 10\%
(cf. Fig.~\ref{completeness}, panel d). The contamination of the
recovered period distribution by incorrect periods is generally small
($\leq$10\%), except in the period range around 1 day and from 5 to 11
days. The excess of periods around 1d mainly results from aliases of
longer input periods (7-17d), while the excess of periods in the range
5-11d arises partly from harmonics of longer input periods (12-20d)
(cf. Fig.~\ref{completeness}, lower panels). This illustrates the fact
that the sampling of the light curves is not optimal for the detection
of long periods, mostly due to a large gap of about 20 nights in the
observing window, between the Maidanak and CFHT datasets
(cf. Figure~\ref{sampling}). Finally, we note that detected periods
with a FAP$\leq$0.01 have a  recovery rate of 95\%, while those with
0.01$<$FAP$\leq$0.05 have a recovery rate of 74\%
(cf. Fig.~\ref{completeness}, lower panels).

The simulations we ran include very small amplitudes and long periods,
none of which may be characteristic of such young stars as the
13~Myr-old h~Per members. Therefore, the actual period detection rate
for h~Per members is likely to be significantly larger than what these
simulations suggest. To illustrate this, restricting the completeness
analysis to input light curves with amplitudes larger than 0.01~mag
raises the period detection rate from 51\% to 71\%, and brings the
recovery rate to 90\% or more for all periods smaller than
10d. Indeed, what is most important here is not so much to reach a
high completeness level but rather to be able to measure reliable
periods in an unbiased way over the whole period range between 0.2d
and at least 10d. The Monte Carlo simulations presented here suggest
that the combination of Maidanak and CFHT datasets allows us to derive
such a relatively robust and largely unbiased rotational period
distribution for h~Per members over this period range.

\subsection{Rotational contamination by field stars}
\label{sec:control}

\begin{figure}
\centering
\includegraphics[width=0.4\textwidth]{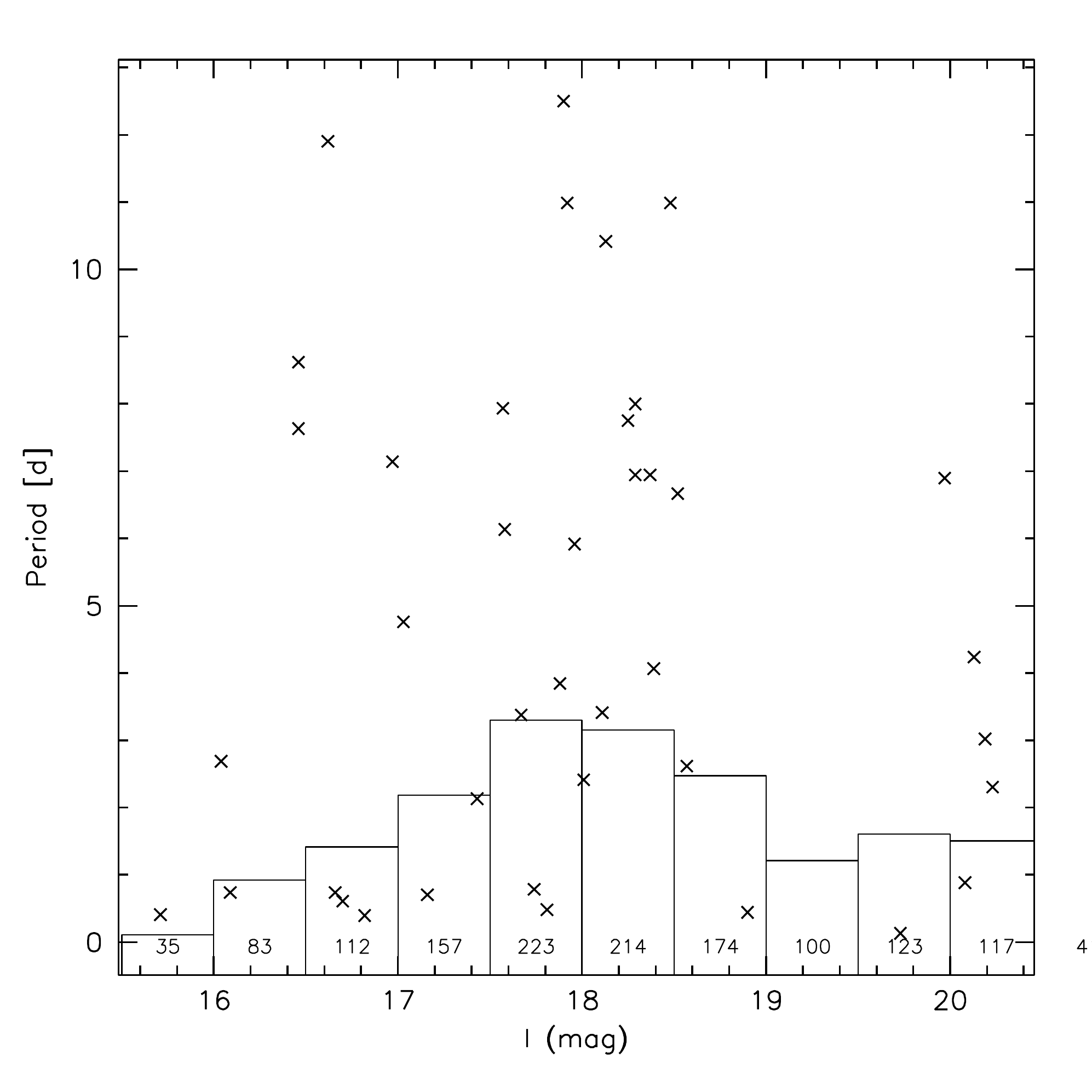}
\caption{The distribution of 40 periods detected in a sample of 1347
  field stars is shown as a function of $i'_{CFHT}$-band magnitude
  (crosses). The lower histogram shows the magnitude distribution of
  the field star sample. The number of field stars is indicated in
  each bin.}
\label{field}
\end{figure}
 
The sample of 2287 h Per photometric candidate members we selected
for period analysis includes a fraction of contaminating field stars
as estimated above (cf. Section~\ref{sec:sample}). In order
to investigate the pollution of the h~Per rotational distribution by
unrelated periodic field stars, we selected a control sample of
photometric {\it non} members. Among the 45,710 photometric non
members, 3169 have at least 800 data points from CFHT and Maidanak in
their combined light curve. Of these, 1347 have been selected in the
$V, V-i'_{CFHT}$ CMD as lying in a locus parallel to the cluster sequence,
either above its upper envelope up to 0.7 mag redder, or below the
lower envelope down to -0.7 mag bluer.  The control sample thus
encompasses a similar magnitude and color range as the sample of h Per
photometric candidate members. We then applied to the control sample
the same period analysis as we did for the photometric
members. Sorting the results by decreasing periodogram peak power, we
vizualized the phased light curves to decide on periodic and
non-periodic objects. Over 1347 objects, we retained 40 objects as
being periodic. The remaining 1307 non periodic field objects were
used to estimate the FAP levels discussed above
(cf. Sections~\ref{sec:periods}, \ref{sec:completeness}).

The distribution of periods for the control sample is shown in
Figure~\ref{field}. Periods typically range from 0.4 to 12d over the
brightness range 16-21 mag.  The shortest periods (P$\leq$1d) we
detect in the control sample probably relate to synchronized field
binaries and the longer periods of a few days may pertain to the young
stellar population of the galactic field (cf. e.g. Briceno et
al. 1997). In any case, the period detection rate is very low for
field stars, amounting to a mere 40/1347 = 3\%. The photometric
contamination of the sample of 2287 h Per candidates by field stars is
estimated to be 18.7\% (cf. Section~\ref{sec:sample}), i.e., about 428
field stars are included in this sample of candidate members. We
expect to detect a period for 3\% of the contaminants, yielding 13
periodic field stars. We thus estimate the contamination of the h~Per
rotational period distribution by field stars to 13/586, i.e., 2.2\%
over the mass range 0.3-1.6M$_\odot$. Furthermore, since the period
distribution we obtain for field stars appears relatively uniform in
the range 1-13 days, this low level of contamination is unlikely to
introduce any significant bias in the period distribution of~h Per
candidate members. We notice, however, an apparent excess of field
stars with periods shorter than 1d, presumably short period binaries.

\subsection {Rotational bias due to synchronized binaries}
\label{sec:bin}

As synchronized binaries are expected to have a different angular
momentum evolution than detached systems and single stars, we
investigate how the former may impact on the period distribution we
derive for h~Per candidate members. To this aim, we first identify
photometric binaries in our sample of 2287 h~Per candidate members
from their location in the $i'_{CFHT}, i'_{CFHT}-K$ CMD. In this
diagram, the binary sequence is fairly well separated from the single
star sequence and we defined the empirical boundary as follows:
\begin{eqnarray*}
  i'_{CFHT}-K &= & 0.475 \,\, i'_{CFHT} -5.625  \quad \textrm{for}
  \quad i'_{CFHT} <19 \\ 
  i'_{CFHT}-K &= & 0.3 \,\, i'_{CFHT} - 2.3  \quad \textrm{for} \quad
  i'_{CFHT} \ge 19 
\end{eqnarray*}
391 objects are redder than this limit and have been classified as
binaries. 

\begin{figure}
\centering
\includegraphics[width=0.5\textwidth]{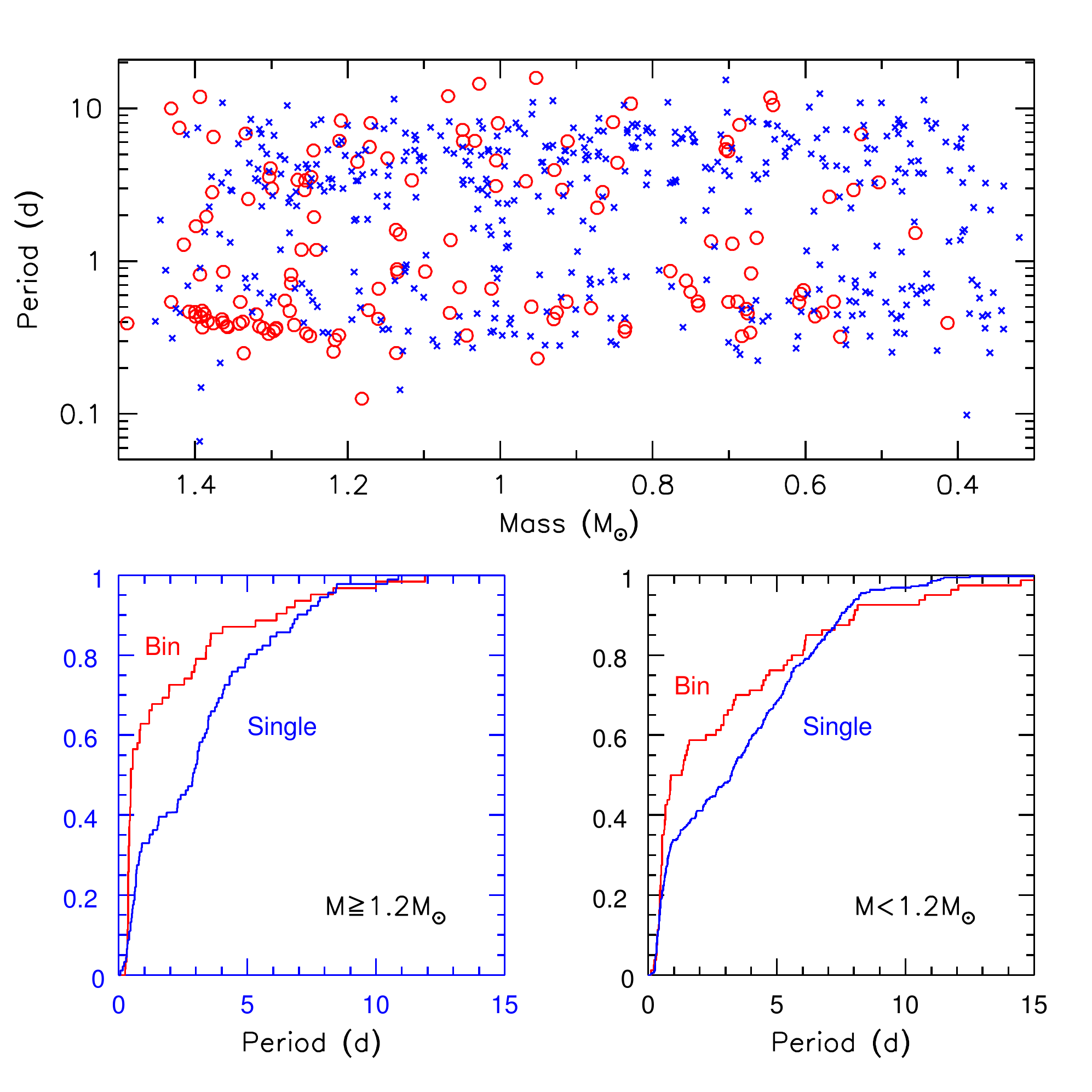}
\caption{{\it Top panel :} The distribution of periods is shown as a
  function of stellar mass for h~Per {\it photometric binary} candidates
  (red circles) and for h Per candidate members lying on the cluster's
  single star photometric sequence (blue crosses). {\it Bottom panels
    :} The cumulative distribution of periods for photometric binaries
  and single stars in 2 mass ranges ({\it left:} M$\geq$1.2M$_\odot$;
  {\it right:} M$<$1.2M$_\odot$). In the 2 mass bins, there is a
  clear excess of short period binaries compared to short period
  single stars (see text). }
\label{binaries}
\end{figure}

Among the 586 periodic h~Per candidate members, 142 are identified as
photometric binaries. The distribution of periods as a function of
mass is shown in Figure~\ref{binaries} for the subsamples of
photometric binaries and of single stars. There is a clear excess of
short period binary systems over the whole mass range from 0.3 to
1.5~M$_\odot$, which is particularly conspicuous in the mass range
above 1.2~M$_\odot$. The cumulative period distribution of single and
binary stars is shown in Fig.~\ref{binaries} in 2 mass bins, on either
side of 1.2~M$_\odot$. A Kolmogorov-Smirnov test indicates that the
probability for the binary and single star period distributions to be
drawn from the same distribution is 9.10$^{-6}$ in the higher mass bin
and 6.10$^{-3}$ in the lower mass one. Clearly, the excess of short
period binaries strongly skews the overall rotational distribution of
the h~Per candidate members.

Indeed, when removing rapid rotators with $P<$1d from the period
distribution, a KS test on the remaining single and binary period
distributions indicate they are the same with a probability of 0.77 in
the higher mass bin and 0.58 in the lower mass one. This suggests
that, unlike synchronized binaries that have a specific rotational
evolution due to tidal effects, the components (or at least the
primary) of wider binary systems appear to undergo a similar angular
momentum evolution as single stars. A similar result was reported for
binaries in the Pleiades cluster (Bouvier et al. 1997).

In order to avoid the bias introduced by synchronized binaries, we
thus remove the 79 photometric binaries detected with P$<1$d from the
sample of 586 periodic h~Per candidate members. Since the period
distribution of wider binary systems appears to be statistically
similar to that of single stars, we keep them for the following
analysis. This leaves us a sample of 508 h~Per candidate members with
measured periods over the mass range from 0.3 to 1.5M$_\odot$, which
we can use to investigate stellar rotation at 13 Myr and angular
momentum evolution prior to the arrival on the main sequence.

\subsection{Period distribution}
\label{sec:distribution}

\begin{figure}
\centering
\includegraphics[width=0.45\textwidth ]{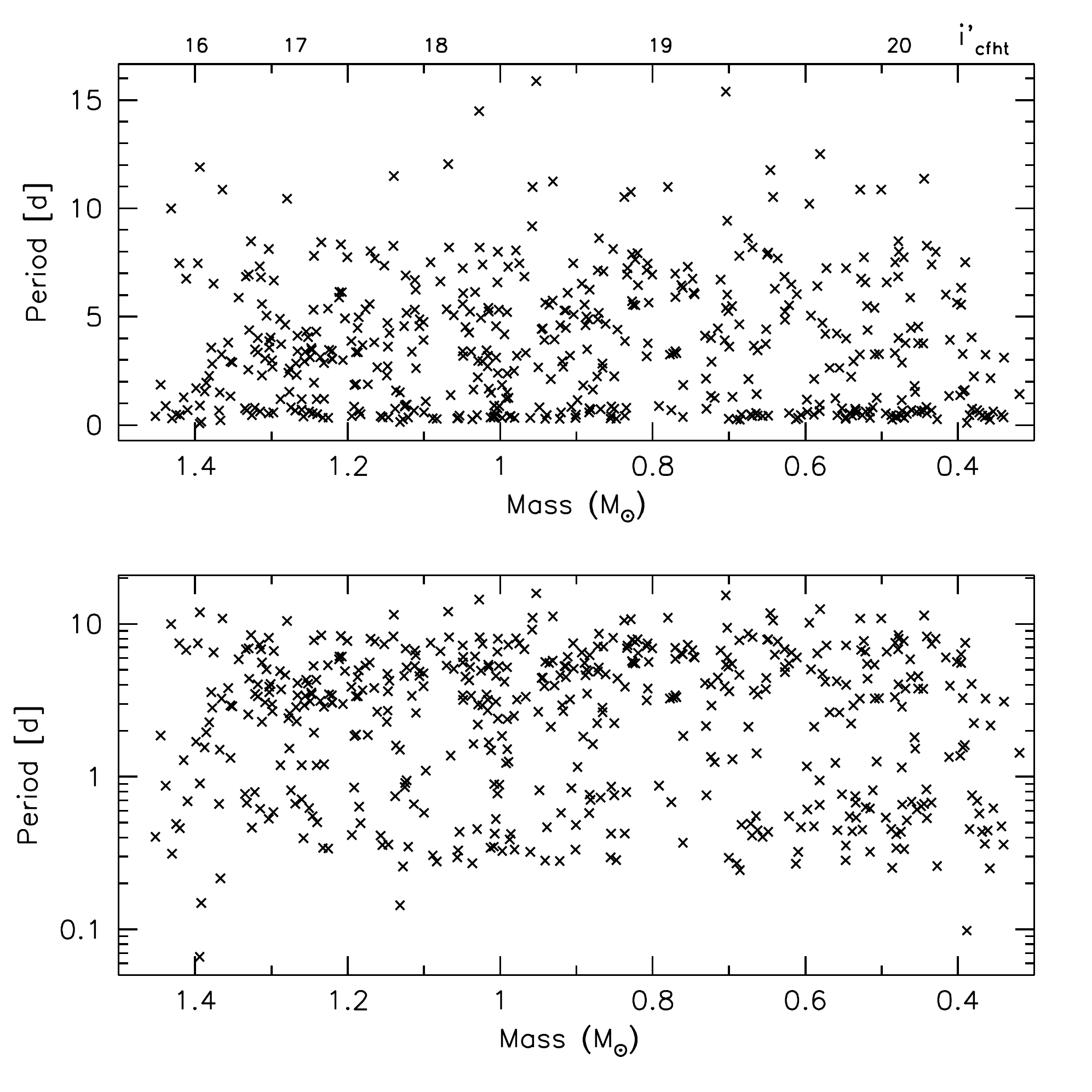}
\caption{The periods detected for 508 h~Per photometric candidate
  members (excluding synchronized binaries with period $\le$1d) are
  shown as a function of stellar mass. Periods are plotted on a linear
  scale in the upper panel, and on a log scale in the lower one. On
  the top panel, the upper x-axis scale corresponds to
  $i'_{CFHT}$-band magnitude.}
\label{permass}
\end{figure}

\begin{figure}
\centering
 \includegraphics[width=0.45\textwidth]{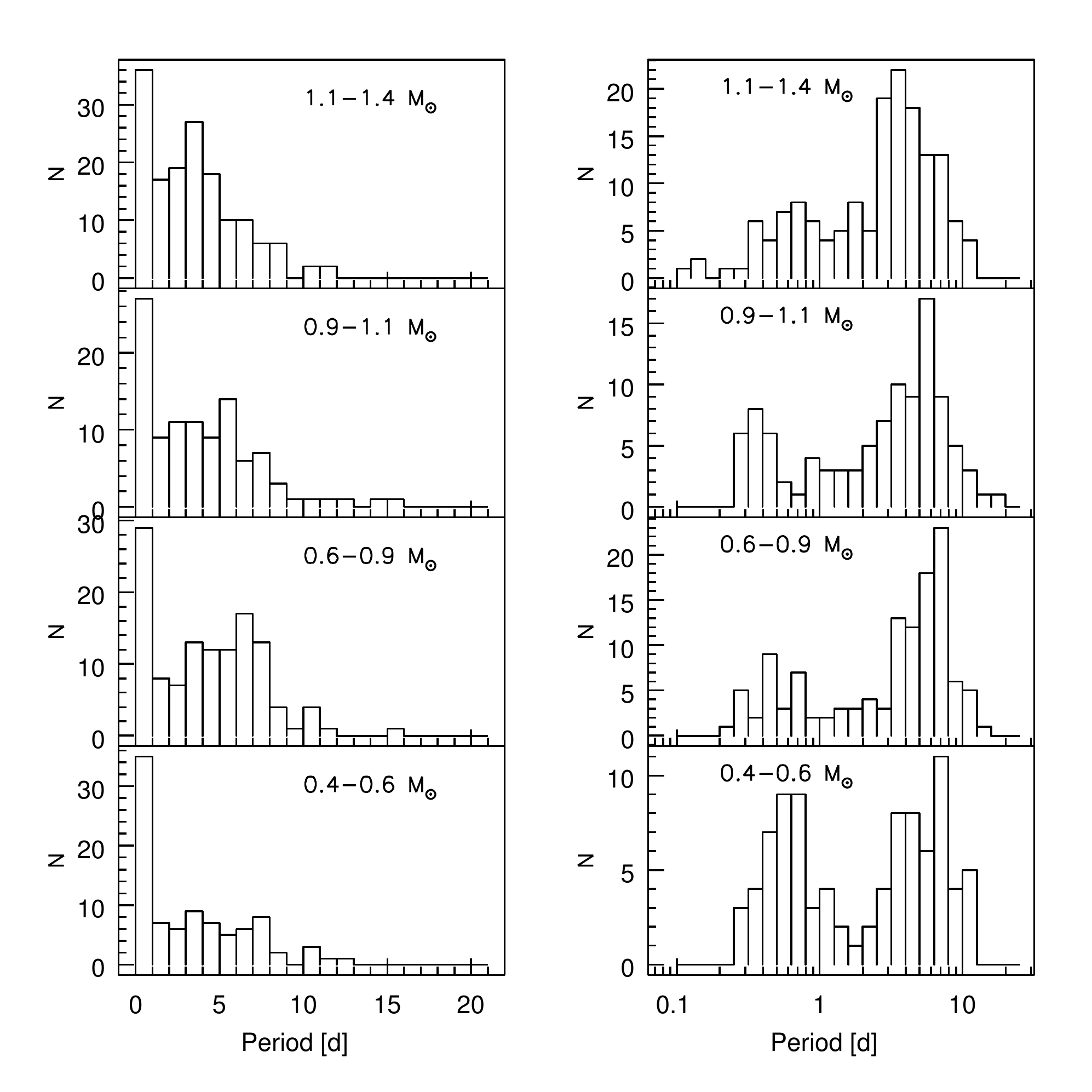}
 \caption{Histograms of the periods of h~Per candidate members in 4
   mass bins. {\it From bottom to top: } 0.4-0.6~M$_\odot$,
   0.6-0.9~M$_\odot$, 0.9-1.1~M$_\odot$, and 1.1-1.4~M$_\odot$. The
   mass bins contain 90, 123, 103, and 152 periodic h~Per candidate
   members, respectively. The histograms are shown on a linear scale on
   the left panels, and on a logarithmic scale on the right panels. }
 \label{perhist}
\end{figure}

The period distribution of h~Per candidate members is shown as a
function of mass in Fig.~\ref{permass}. It extends widely from
$\sim$0.2 to $\sim$12 days at all masses, from $\sim$0.3 to
$\sim$1.4~M$_\odot$, and its lower and upper envelopes do not show any
clear dependence with mass.  We have divided our sample of periodic
objects in 4 mass bins, centered on 0.5, 0.75, 1.0 and 1.25~M$_\odot$,
respectively, and plot their distributions in Fig.~\ref{perhist}. In
each mass bin, most of the periods lie in the range 2-10~d, with a
tail of slower rotators extending up to about 20 days, and a peak of
fast rotators, with periods less than 1d. This peak is particularly
conspicuous at the lowest masses and becomes weaker in the highest
mass bin. The fraction of objects rotating faster than 2 days is
$47\pm8$\%, $31\pm6$\%, $35\pm7$\% and $34\pm5$\% from the lowest to
the highest mass bin. On a log scale, the period distributions in each
mass bin appear somewhat bimodal, especially at low masses again, with
a primary peak around 8-10 days, and a secondary peak at a fraction of
a day. Such bimodal period distributions, albeit on a linear period
scale, have previously been reported for young stars in the ONC at
1Myr, with peaks at 7-9 days and 1-2 days (e.g., Herbst et al. 2001).

Kolmogorov-Smirnov tests were run to compare the period distribution
of each mass bin to the next (i.e. 0.4-0.6~M$_\odot$ vs
0.6-0.9~M$_\odot$, 0.6-0.9~M$_\odot$ vs 0.9-1.1~M$_\odot$ and
0.9-1.1~M$_\odot$ vs 1.1-1.4~M$_\odot$). The probability that the
distributions are drawn from the same overall parental distribution is
0.04, 0.17 and 0.33, respectively.  Hence, while the period
distribution does not seem to depend on mass at an age of 13~Myr above
0.6 M$_\odot$, there is a hint for the 0.4-0.6~M$_\odot$ mass objects
to rotate faster in average. This is in agreement with the finding of
previous studies indicating a larger fraction of fast rotators among
the lowest mass stars (e.g. Herbst et al. 2001, Irwin et al. 2007b,
2008a, 2008b).

\begin{figure}
\centering
\includegraphics[width=0.45\textwidth ]{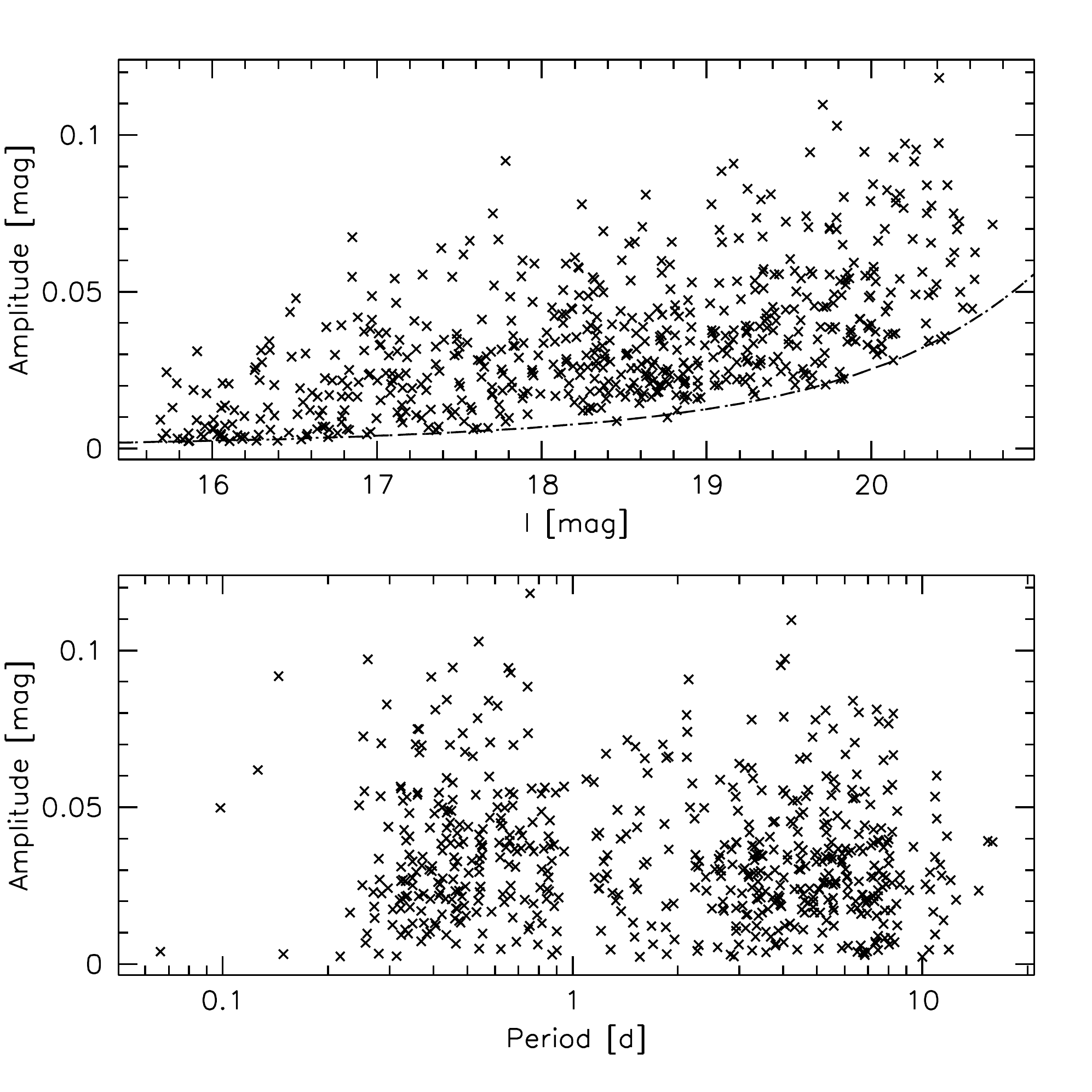}
\caption{The photometric amplitudes derived from a sinusoidal fit to
  the periodic light curves are shown as a function of $i'_{CFHT}$
  magnitude (upper panel) and period (lower panel).}
\label{amp}
\end{figure}

The photometric amplitudes of the 508 periodic h~Per candidate members
are plotted as a function of $i'_{CFHT}$-band magnitude and period in
Fig.~\ref{amp}. The lower envelope of the photometric amplitude is
seen to increase with $i'_{CFHT}$-band magnitude. This trend clearly
points to an observational bias, as it is increasingly difficult to
detect lower amplitude variables among fainter objects
(cf. Fig.~\ref{rms}). The upper envelope of the photometric amplitude,
however, also appears to increase towards fainter, i.e., lower mass,
objects. This suggests that lower mass stars either have a larger
fractional coverage by stellar spots, or that the spot distribution at
their surface is more asymmetric than in higher mass stars, thus
leading to a larger contrast along the rotational cycle. No
correlation is found betweeen photometric amplitude and rotational
period in Fig.~\ref{amp}. A K-S test on photometric amplitudes for
stars with periods shorter and longer than 1d returns a probability of
0.06 of being drawn from the same distribution. Hence, a weak trend,
if any, might be present with median values of 0.034 and 0.029~mag for
the photometric amplitudes of fast and slow rotators,
respectively. Taken together, these results suggest that the origin of
increasing maximum photometric amplitude towards lower mass stars at
13~Myr is not primarily linked to rotation but more likely related to
the internal structure of the pre-main sequence objects, and possibly
to the extent of their convective zones.

\section{Discussion}

\subsection {Angular momentum distribution at 13 Myr}

\begin{figure}
\centering
 \includegraphics[width=0.45\textwidth]{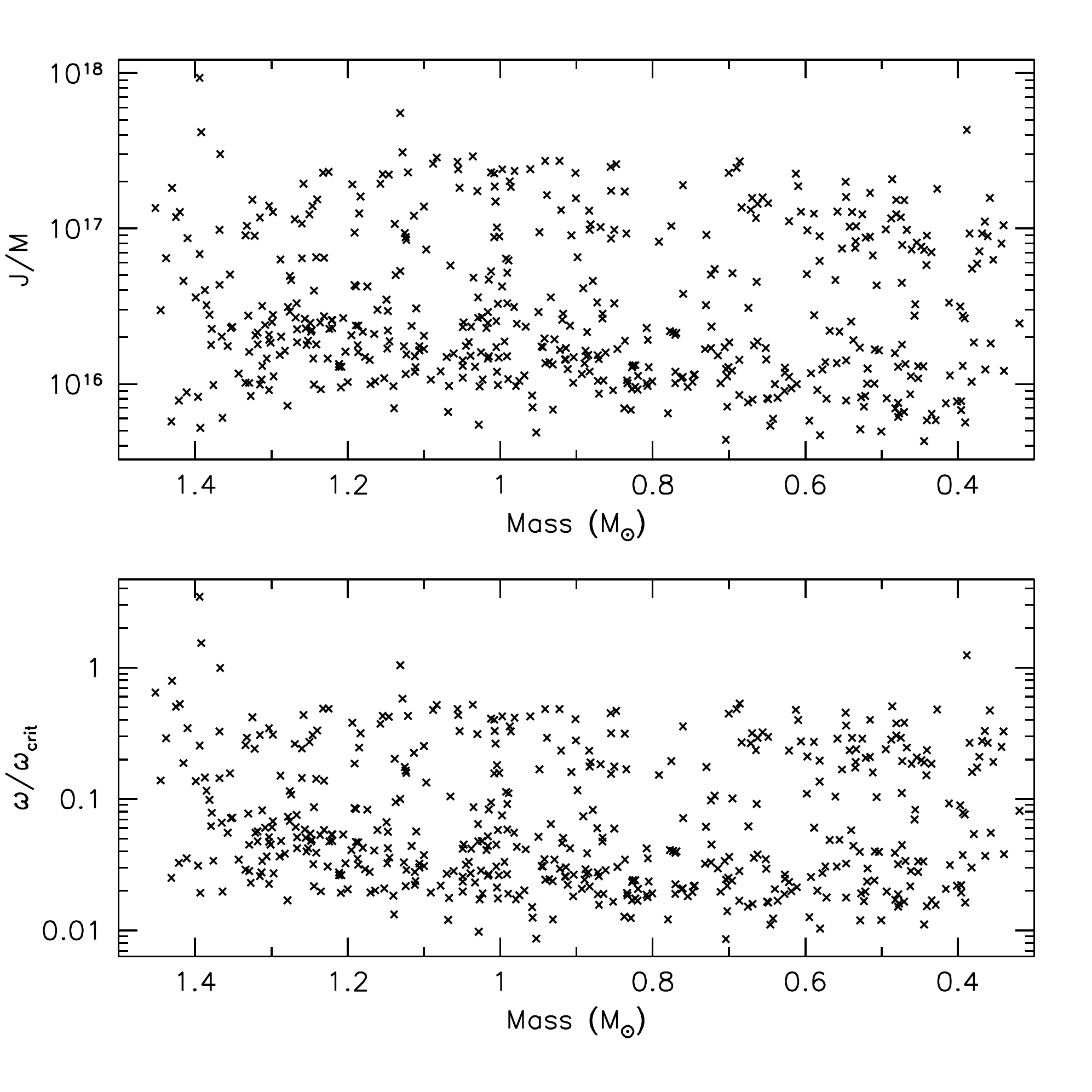}
 \caption{The specific angular momentum ($cm^2 s^{-1}$) of 508 periodic h~Per
   candidate members in shown as a function of mass in the upper
   panel. The ratio of their angular velocity to the break-up velocity
   ($\omega_{crit} = (GM)^{1/2}R^{-3/2}$) is shown as a function of
     mass in the lower panel. }
  \label{breakup}
\end{figure}

At 13 Myr, the accretion process has terminated and the stars are no
longer magnetically coupled to their disk (Currie et al. 2007a; Fedele
et al. 2010). Measurements of rotational periods in the h Per cluster
thus provide the initial angular momentum distribution of freely
evolving stars. We computed the specific angular momentum of the 508
stars with known rotational periods assuming uniform rotation, i.e.,
$J/M = I\cdot \Omega/M$, where the stellar moment of inertia $I$ is
derived at each mass from Baraffe et al. (1998) 13~Myr models, and
$\Omega=2\pi/P_{rot}$. The resulting distribution of specific stellar
angular momenta at 13~Myr is shown as a function of mass in
Figure~\ref{breakup}. A large dispersion is observed, between
$\log(J/M) \sim 15.7-17.5$, with a weak trend for increasing
$J/M$ with mass. The lower values of the specific angular momentum of
solar-mass stars at 13~Myr ($(J/M)_{min} \sim 10^{16}$ cm$^2$s$^{-1}$)
are still about an order of magnitude larger than that of the present-day
Sun ($J/M = 9.2~10^{14}$ cm$^2$s$^{-1}$, Pinto et al. 2011).

Rebull (2001) and Herbst \& Mundt (2005) reported the distribution of
specific angular momenta for low mass stars in the Orion Nebula
Cluster and its flanking fields at an age of $\sim$1~Myr. They both
find a large dispersion of $J/M$, with a range of $\log(J/M) \sim
16.0-17.7$ and a peak at $\log(J/M)\sim$16.5 in Rebull (2001)
\footnote{Rebull (2001) computed angular momenta assuming a gyration
  factor $k^2$, where $I=k^2 M R^2$, of 2/5 corresponding to a solid
  body rotating sphere of uniform density. However, for their sample
  of fully convective stars, the correct gyration factor is 0.205
  (polytrope n=1.5, Horedt 2004). We therefore reduced Rebull's $J/M$
  value by a factor of 2 to compare with ours.}, and a range of
$\log(J/M) \sim 16.2-17.5$ with a peak at $log(J/M)\sim$16.5 in Herbst
\& Mundt (2005)\footnote{Herbst \& Mundt (2005) computed the
  specific angular momentum of the surface shell of ONC stars, using a
  gyration factor $k^2\sim 2/3$. For the whole star, however, assuming
  it is fully convective, rotates as a solid body and neglecting
  rotational distortion, $k^2=0.205$. To compare with our study, we
  therefore reduced Herbst \& Mundt (2005) specific angular momentum
  values by a factor 3.3, which is appropriate for all but the most
  rapid rotators.}. The scatter in $J/M$ in both studies (1.7 and
1.3~dex, respectively) is roughly similar to what we derive (1.8~dex)
for the $J/M$ distribution of h~Per low-mass members at 13~Myr.
However, the peak of the $J/M$ distribution for h~Per members
($\log(J/M)\sim$16.2) appears to be slighty shifted towards lower
values. The decrease of the specific angular momentum of low-mass
stars between $\sim$1 and 13~Myr presumably reflects angular momentum
loss due to the star-disk interaction during the early PMS (Rebull et
al. 2004).
  
Figure~\ref{breakup} also shows the ratio of angular velocity to
critical velocity, $\omega/\omega_{crit} $, as a function of mass for
the 508 periodic h~Per candidate members. For a few stars (including
HPer-215 and HPer-513; see Section~\ref{sec:periods}), this ratio
exceeds unity, suggesting the photometric period could reflect either
pulsation or orbital motion instead of rotation. The distribution is
strongly peaked at $\omega \leq 0.1 \omega_{crit}$, with a median
value of 0.048, and a tail extending up to $\omega \sim 0.5
\omega_{crit}$. Of the 508 periodic members, 492 (97\%) have $\omega
\leq 0.5 \omega_{crit}$. No evidence for a mass dependence is found,
except perhaps at masses larger than 1.1~M$_\odot$ where the lower
envelope of the $\omega/\omega_{crit}$ distribution may slightly
increase . The range and shape of the $\omega/\omega_{crit}$
distribution we derive for low-mass stars at 13~Myr is qualitatively
similar to that reported for similar stars in the ONC cluster at an
age of 1~Myr (Stassun et al. 2001). When projected to the ZAMS
assuming no angular momentum loss, i.e., $\omega/\omega_{crit}\propto
I^{-1}R^{1.5}$, the $\omega/\omega_{crit}$ ratio increases by a factor
of 1.4 to 2.3 over the mass range 0.4-1.4~M$_\odot$. The projected
distribution on the ZAMS has a median value of $\omega/\omega_{crit}
\sim 0.1$, and 80 stars (16\%) have $\omega > 0.5 \omega_{crit}$,
including 15 stars with $\omega/\omega_{crit} \geq 1$. This suggests
that some angular momentum loss must occur during the late PMS
evolution, an issue to which we return in the next section.


\subsection {Angular momentum evolution to the ZAMS}

The derivation of hundreds of rotational periods for low-mass stars in
the 13~Myr-old h~Per cluster provides a new time step to investigate
pre-main sequence angular momentum evolution. This time step, not
previously covered by other cluster studies (e.g. Irwin \& Bouvier
2009; Messina et al. 2010, 2011) is particularly interesting as it
marks the end of the PMS disk accretion process. Disk accretion is
thought to be largely terminated by 10~Myr (Kennedy \& Kenyon 2009;
Fedele et al. 2010), leaving at most a few percent of stars still
actively accreting from their disk in h~Per (Currie et al. 2007c). The
so-called ``disk-locking process'' by which stars are prevented from
spinning up during their early PMS evolution is therefore over for
most of the h~Per low mass population. These stars are thus expected
to freely spin-up as they contract towards the zero-age main sequence
(ZAMS) that they eventually reach at an age of 22, 33, 66, and 100~Myr
for a mass of 1.2, 1.0, 0.7, and 0.5~M$_\odot$, respectively. The
rotational distribution of h~Per members is thus particularly suited
to investigate stellar spin-up at the end of PMS evolution and on the
approach to the ZAMS.

\begin{figure}
\centering
 \includegraphics[width=0.45\textwidth]{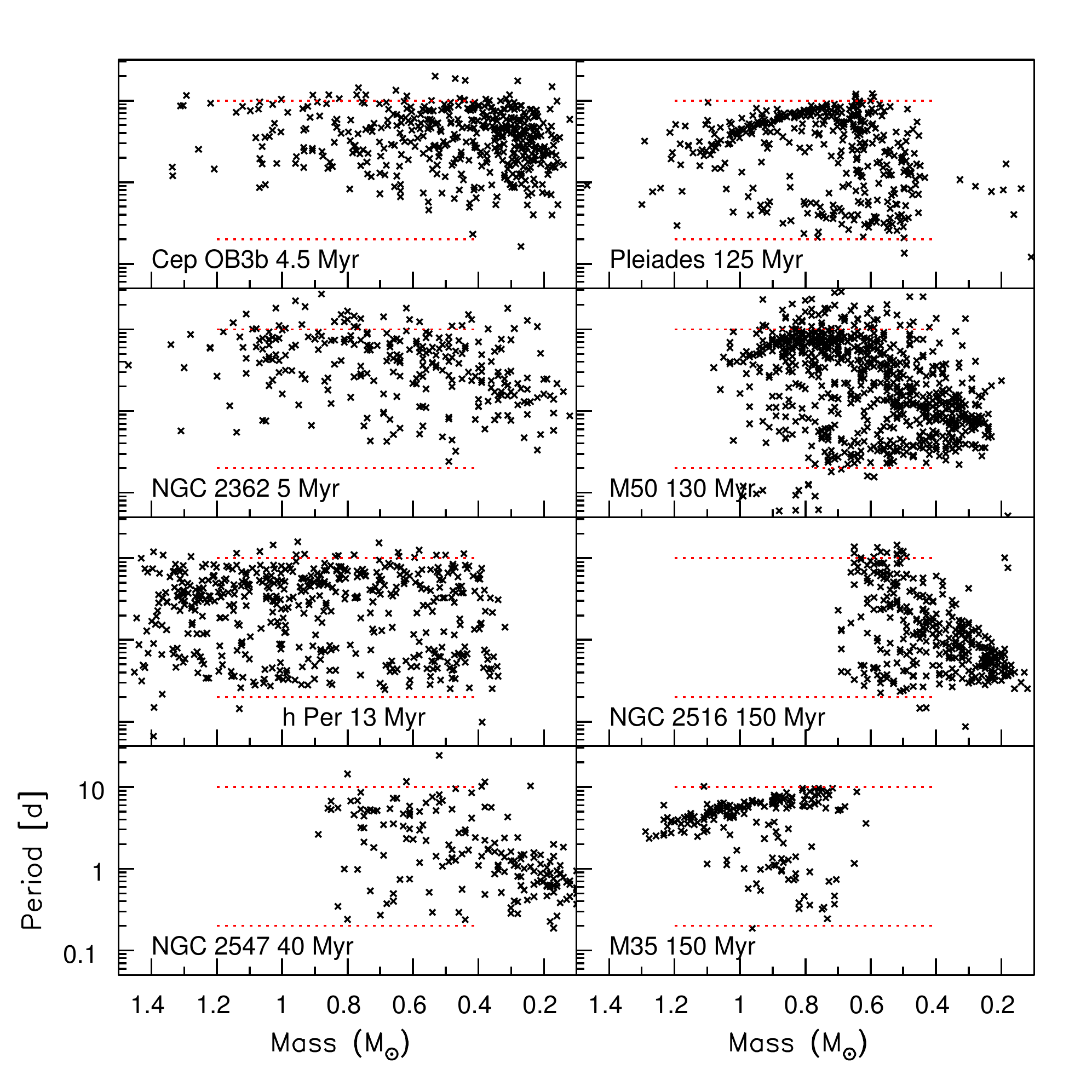}
 \caption{Period versus mass distributions for clusters with an age
   ranging from 4 to 150~Myr. The clusters' name and age are given in each
   panel.  Red dotted lines are drawn at periods of 0.2 and 10 days to
   guide the eye. References: Cep~OB3b: Littlefair et al. (2010);
   NGC~2362: Irwin et al. (2008a); NGC~2547: Irwin et al. (2008b);
   Pleiades: Hartman et al. (2010); M~50: Irwin et al. (2009); NGC~2516:
   Irwin et al. (2007b); M35: Meibom et al. (2009). }
  \label{permassall}
\end{figure}

Figure~\ref{permassall} compares the rotational distribution of h~Per
members with those of solar-type and lower mass members of various
open clusters over the age range from 5 to 150 Myr. At all ages, a
significant scatter is seen in the rotational period distributions,
which cannot be accounted for by observational errors, but reflects a
true dispersion of rotation rates during the PMS. In these period-mass
plots, both early-PMS and ZAMS clusters exhibit some similarities with
h~Per but also striking differences. The upper envelope of the period
distribution, located at $\sim$10~day in h~Per, does not appear to
evolve much between 5 and 40~Myr over the mass range from 0.4 to
0.9~M$_\odot$. This suggests that at least a fraction of slow rotators
 are prevented from spinning up over this timescale. By the age
of the Pleiades (125~Myr) the upper envelope of the distribution
has decreased towards faster rotation for masses larger than
0.7~M$_\odot$. A significant fraction of lower mass stars, even at
this age, still exhibit long periods, suggesting that the pre-ZAMS
spin up is more efficient for solar-mass than for low mass stars.

In contrast, the lower envelope of the period distribution exhibits
quite a drastic evolution from 5~Myr to the ZAMS. In h~Per, the lower
envelope appears rather flat over the mass range 0.4-1.1~M$_\odot$, at
a period of $\sim$0.2-0.3d. In younger clusters, the minimum period
seems to strongly depend on mass, ranging from $\sim$0.4-0.5d at
0.4-0.6~M$_\odot$ to 1.0d or more for solar-type stars. This provides
good evidence for PMS spin up for the fast rotators between 5 and
13~Myr.  While the mass sampling and period statistics is
unfortunately sparser in NGC2547, its 40~Myr-old rotational
distribution does not show any clear evolution for the fastest
rotators between 13 and 40~Myr over the mass range
0.4-0.8~M$_\odot$. From 13 to 130~Myr, however, the fastest rotators
in the 0.8-1.1~M$_\odot$ mass range have been spun down, presumably
upon their arrival on the ZAMS, while lower mass stars down to
0.4~M$_\odot$ still exhibit the same maximum rotation rate of
$\sim$0.3d. The most striking difference between the 13~Myr-old h~Per
cluster and ZAMS clusters at an age of 125-150~Myr, is the lack of
mass-dependency in the rotational distribution of the former while the
latter exhibit a narrow rotation-mass relationship for masses larger
than about 0.7~M$_\odot$. Indeed, the largest scatter of rotation
rates over the investigated mass range is observed for the 13~Myr-old
h Per cluster, at the end of the PMS accretion phase, a result that
supports the role of disk accretion in establishing the initial
dispersion of stellar angular momentum in low-mass stars (Bouvier et
al. 1993; Rebull et al. 2004).

\begin{figure}
\centering
 \includegraphics[width=0.45\textwidth]{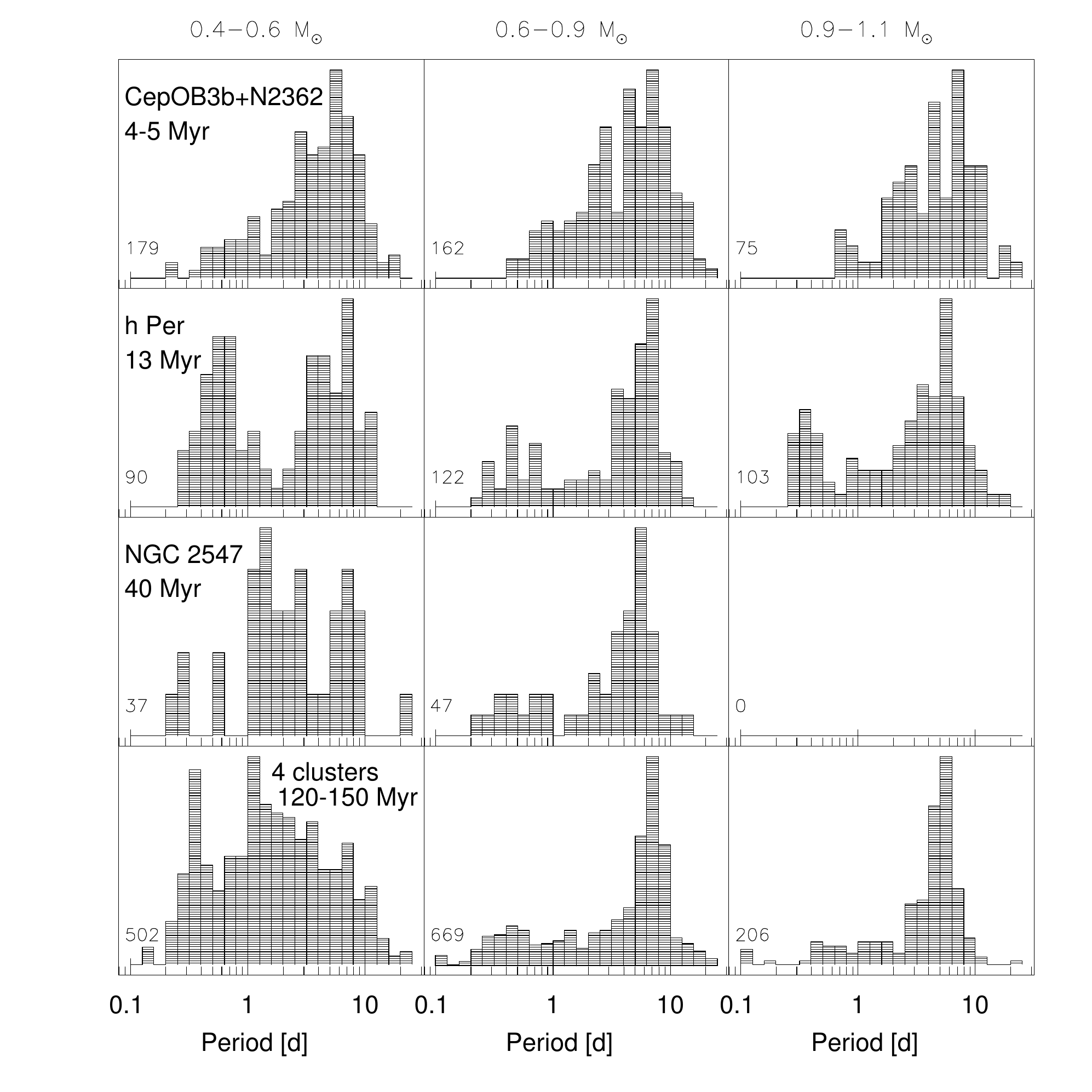}
 \caption{Histograms of period distributions in 3 mass bins for clusters with an age
   ranging from 4 to 150~Myr. The clusters' name and age are given in each
   panel.   {\it Left:} 0.4-0.6~M$_\odot$, {\it Middle:}
   0.6-0.9~M$_\odot$, {\it Right:} 0.9-1.1~M$_\odot$. The number of
   stars in each mass bin is indicated in the lower left corner of the
 panels.}
  \label{perhistall}
\end{figure}

\begin{table}
\label{inert}
\center
\caption {\label{inertia} Stellar moment of inertia  ($I/I_\odot$) }
\begin{tabular}{lllll}
\hline
Age (Myr) &	0.5~M$_\odot$	 &     0.75~M$_\odot$	&  1.0~M$_\odot$		&  1.25~M$_\odot$\\
\hline
5 & 1.58 & 3.40 & 5.61 & 8.49 \\
13 & 0.81 & 1.74 & 2.89 & 3.85 \\
40 & 0.37 & 0.74 & 1.00 & 1.39 \\
130 & 0.22 & 0.55 & 0.95 & 1.34 \\
\hline
\end{tabular}
\end {table}

A more detailed comparison of the rotational distributions at 4 time
steps between the early PMS and the ZAMS is provided in
Fig.~\ref{perhistall} for 3 mass bins centered at 0.5, 0.75, and
1.0~M$_\odot$, respectively. We do not consider here stars in the mass
range 1.1-1.4~M$_\odot$ as their rotational period distributions are
still missing for most young open clusters. In order to improve
statistics, and to smooth out any dependence of rotation on
environment (cf. Littlefair et al. 2010), the two 4-5~Myr clusters
have been merged, as have been the 4 clusters in the age range
125-150~Myr.  The temporal evolution of these distributions can now be
compared to the expected PMS spin up in the absence of angular
momentum loss.  Table~\ref{inert} lists the evolution of the stellar
moment of inertia resulting from the star's contraction and the
development of a radiative core. In all 3 mass bins considered here,
the reduction of the moment of inertia would translate into a spin up
by a factor of $\sim$2 from 5 to 13~Myr, and by a factor of $\sim$2.5
from 13 to 40~Myr. The observed distributions do suggest some PMS spin
up between 5 and 13~Myr in all mass bins, with the bulk of fast
rotators (P$<$2d) shifting towards shorter periods, as expected from
the evolution of the stellar structure. The evolution of the fast
rotators is less clear from 13 to 40~Myr. This is partly due to the
small statistics affecting the distribution of NGC~2547 in the
0.4-0.6~M$_\odot$ mass range, although in the better sampled
0.6-0.9M$_\odot$ range, little evolution is observed for fast
rotators, contrary to model expectations. Unfortunately, no data exist
for NGC~2547 solar-type stars and only a few rotational periods have
been published in the 0.9-1.1$M~_\odot$ range (Messina et al. 2011).
Slow rotators (P$\geq$2d) as a whole do not appear to spin up as much
as fast ones. In the 0.6-0.9$M~_\odot$ range, the peak of the slow
rotator distribution varies from $\sim$7.5d at 5~Myr, to $\sim$6.5d at
13~Myr, and $\sim$5.5d at 40~Myr, i.e., much less than the reduction
of the stellar moment of inertia would imply. Similar conclusions are
reached for the other mass bins. Clearly, some angular momentum loss
must occur in at least a fraction of slow rotators over the 5-40~Myr
age range to compensate for the spin up due to contraction.

While the rotational distributions of fast and slow rotators provide
some clues to the angular momentum evolution of PMS stars, the
complete distributions have to be modelled to understand the processes
at play. We therefore computed angular momentum evolution models
starting from the observed 5~Myr rotational distributions as initial
conditions and evolved them at ages of 13, 40 and 130~Myr to compare
with observations. The models assume all stars are released from their
disk at 5~Myr, therefore does not include any ``disk locking'' process
after this age. Angular momentum loss due to stellar winds are
included in the way described in Irwin \& Bouvier
(2009). Core-envelope decoupling is also included in the model through
the introduction of a coupling timescale $\tau_c$ over which angular
momentum is exchanged between the radiative core and the convective
envelope (Allain 1998). Previous modelling efforts suggest that
$\tau_c$ is significantly shorter for rapid rotators than for slow
ones (Bouvier 1997; Irwin et al. 2007b; Bouvier 2008; Irwin \& Bouvier
2009; Denissenkov et al. 2010; Spada et al. 2011). We illustrate here
3 classes of models: solid-body rotation models ($\tau_c$=1~Myr),
decoupled models with a constant coupling timescale
($\tau_c$=50-100~Myr), and a velocity-dependent coupling model, with
$\tau_c = \tau_{c0}\cdot(\omega/\omega_\odot)^{-1}$~Myr and
$\tau_c0$=100-500~Myr.  The other parameters of the models, namely the
angular momentum loss scaling and saturation velocity (see Irwin \&
Bouvier 2009) are summarized in Table~\ref{mod}. Owing to the observed
scatter of rotation rates among field stars (Harrison et al. 2012),
the models are required to reproduce the angular velocity of the Sun
to within 30\% for solar-type stars, and up to a factor of 2 lower
rotation rates for lower mass stars. In addition, the solar-mass
models must comply with heliosismology results that indicate no
residual excess rotation in the radiative core of solar-mass stars by
the age of the Sun (Thompson et al. 2003).  Stellar evolution models
for 0.5, 0.8, and 1.0~M$_\odot$ stars are taken from Baraffe et
al. (1998).

\begin{table}
\caption{Angular momentum model parameters}
 \begin{tabular}{c|cc|c}
Mass bin & \multicolumn{2}{c|}{Wind braking$^\dagger$} & Core-envelope \\
{\it (Stellar model)}&&&coupling\\
&&

\\
\hline
$M$ & $K$ & $\omega_{sat}$ & $\tau_c$ \\
($M_\odot$) & ($10^{47} gcm^2s$) & ($\omega_\odot$) & ($yr$) \\ 
    \hline
               & 3.75 & 8.0 & 10$^6$ \\
 0.9-1.1  & 7.5   & 8.0 & 10$^8$ \\
 {\it(1.0)} & 5.6   & 8.0 & 10$^8\cdot(\omega/\omega_\odot)^{-1}$ \\
\hline
               & 3.75 & 6.0 & 10$^6$ \\
 0.6-0.9  & 15   & 6.0 & 5.10$^7$ \\
  {\it(0.8)} & 4.5  & 6.0 & 5.10$^8\cdot(\omega/\omega_\odot)^{-1}$ \\
\hline
                & 2.25 & 4.5 & 10$^6$ \\
 0.4-0.6   & 4.5 & 4.5 & 5.10$^7$ \\
  {\it(0.5)} & 3.75 & 4.5 & 5.10$^8\cdot(\omega/\omega_\odot)^{-1}$ \\
\hline
\multicolumn{4}{l}{} \\
\multicolumn{4}{l}{$^{^\dagger}\left({dJ\over{dt}}\right)_{\rm wind} = \left\{ \begin{array}{r}
 -K\ \omega^3\ \left({R \over{R_\odot}}\right)^{1/2} \left({M \over{M_\odot}}\right)^{-1/2}, \omega < \omega_{sat} \\
 -K\ \omega\ \omega_{sat}^2\ \left({R \over{R_\odot}}\right)^{1/2} \left({M \over{M_\odot}}\right)^{-1/2}, \omega \ge \omega_{sat}
\end{array} \right.$}\\
 \end{tabular}
\label{mod}
\end{table}

\begin{figure}
\centering
\includegraphics[width=0.5\textwidth ]{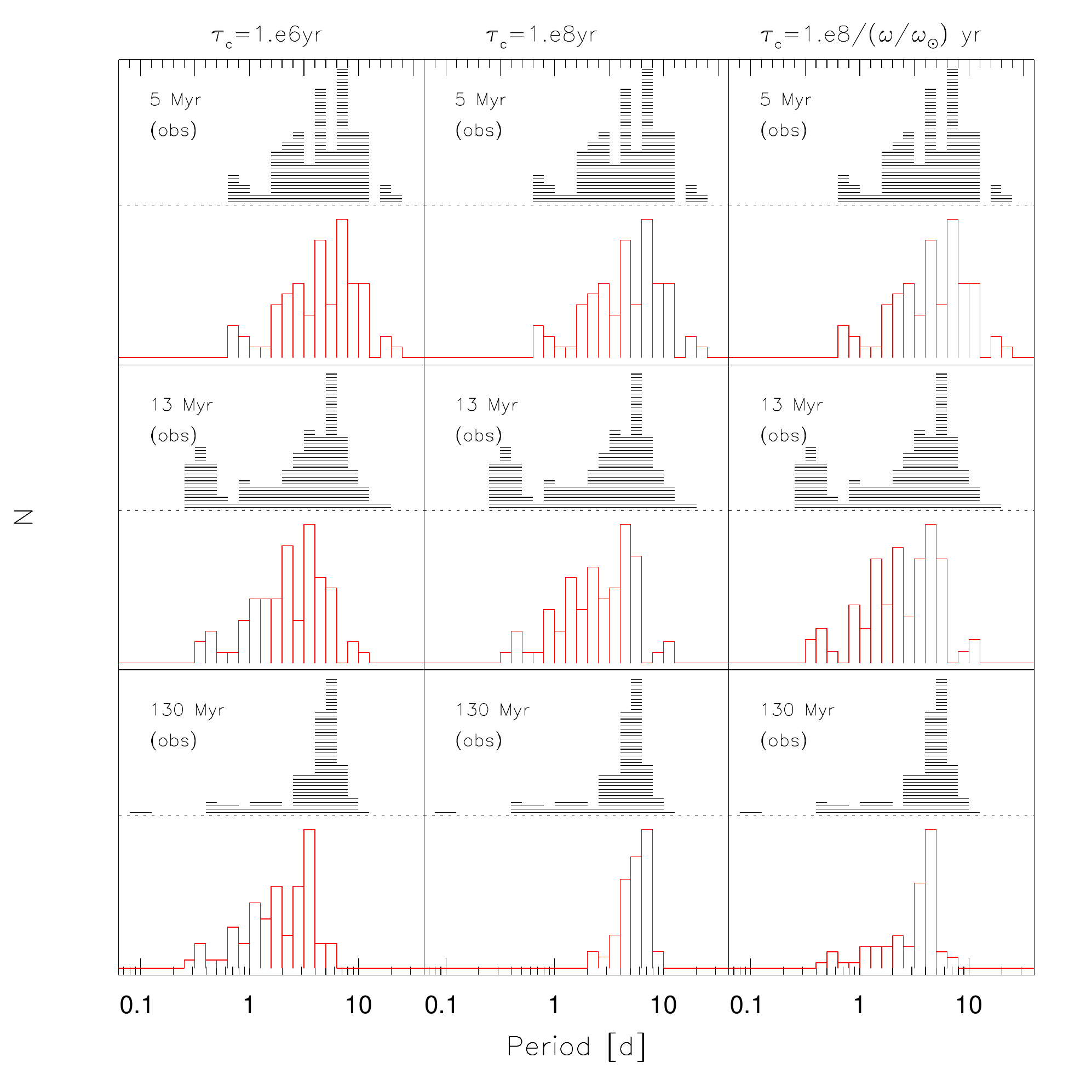}
\caption{The initial distribution of periods at 5~Myr (top panels) for
  0.9-1.1~M$_\odot$ stars is projected to 13~Myr (middle panels) and
  130 Myr (bottom panels) using 3 models : strong core-envelope
  coupling ($\tau_c=1$~Myr, left panels), weak core-envelope coupling
  ($\tau_c=100$~Myr, middle panels), and a velocity-dependent coupling
  timescale ($\tau_c=100\cdot(\omega/\omega_\odot)^{-1}$~Myr, right
  panels). The observed distributions are shown as black shaded
  histograms, the projected ones as red bar histograms.}
\label{pproj1mo}
\end{figure}
 
The results for solar-mass stars are shown in Figure~\ref{pproj1mo}
where the initial period distribution at 5~Myr is projected forward in
time using the 3 models to ages of 13 and 130~Myr, and compared to the
observed distributions at those ages. It is seen that, starting from
5~Myr, the projected distributions at 13~Myr do not depend much of the
assumed model. This is because the radiative core only recently
developed at this stage, and the amount of core-envelope decoupling
hardly impacts on the early PMS rotational evolution. Indeed, the
projected distributions at 13~Myr encompass the full range of
rotational periods observed for solar-mass stars in the h~Per
cluster. These results support the hypothesis that most of the disk
locking process is over in solar-mass stars by 5-10~Myr (Rebull et
al. 2004), which is consistent with current estimates of disk
lifetimes (Fedele et al. 2010). Once on the ZAMS, however, significant
differences occur between the models. Solid-body rotation models
succesfully reproduce the fastest ZAMS rotators, but fail to account
for the bulk of slow rotators in Pleiades-like clusters. On the
opposite, distributions projected with fully decoupled models do yield
a population of ZAMS slow rotators, but do not exhibit the observed
tail of fast rotators.  Only the $\omega$-dependent $\tau_c$ model
($\tau_c=100 \cdot(\omega/\omega_\odot)^{-1}$~Myr) simultaneously accounts for
both the peak of slow rotators and the high-velocity tail observed in
ZAMS clusters. This confirms the need for a core-envelope
coupling timescale that depends upon the rotation rate.  As discussed
in Bouvier (2008), the different PMS rotational history of slow and fast
rotators may leave imprints in the properties of field stars once
their angular velocities have converged on the MS, such as an increased Li
dispersion (e.g.. Pasquini et al. 1997).

\begin{figure}
\centering
\includegraphics[width=0.5\textwidth ]{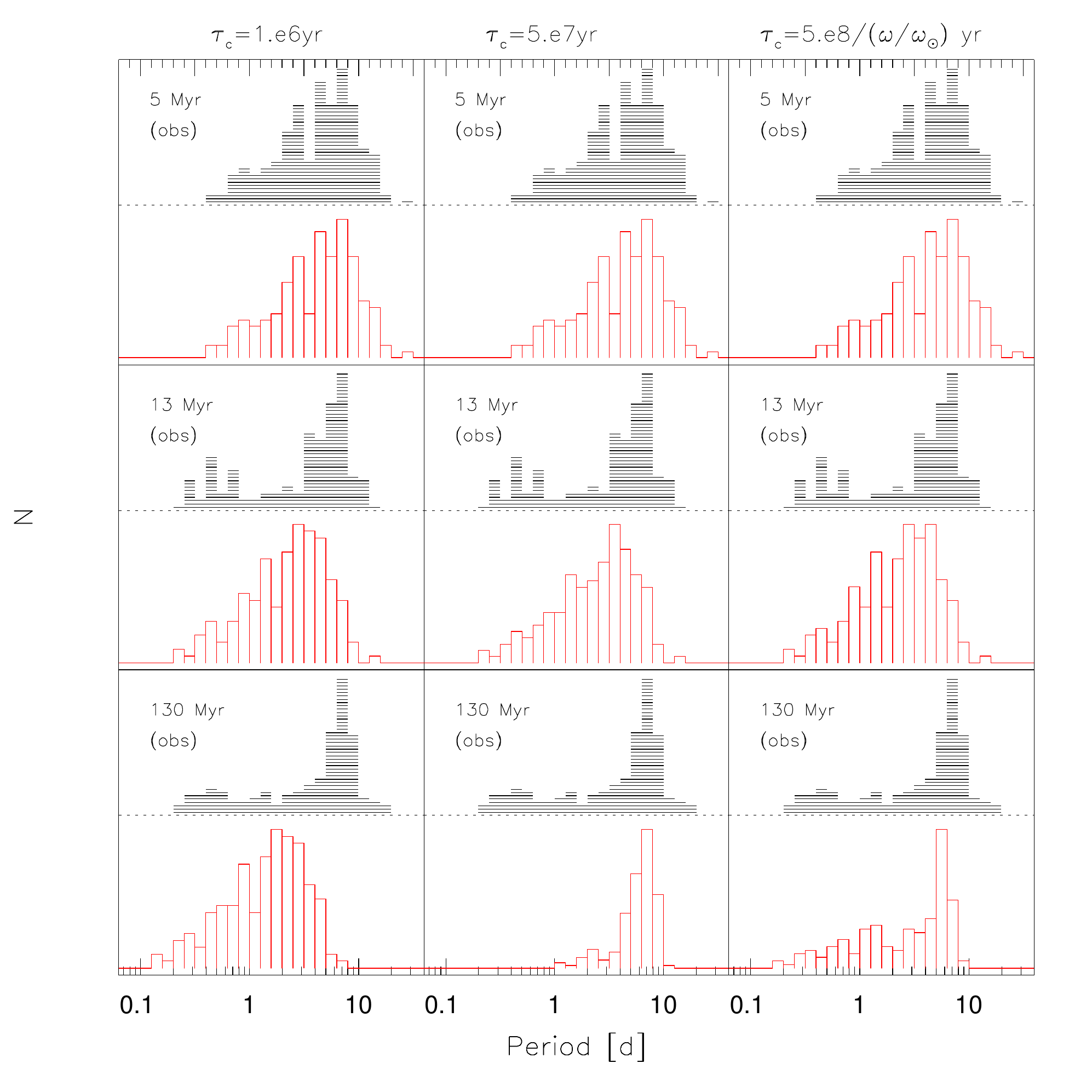}
\caption{ The initial distribution of periods at 5~Myr (top panels)
  for 0.6-0.9~M$_\odot$ stars is projected to 13 and 130 Myr (top to
  bottom panels) using 3 models : strong core-envelope coupling
  ($\tau_c=1$~Myr, left panels), weak core-envelope coupling
  ($\tau_c=50$~Myr, middle panels), and a velocity-dependent coupling
  timescale ($\tau_c=500 \cdot(\omega/\omega_\odot)^{-1}$~Myr, right
  panels). The observed distributions are shown as black shaded
  histograms, the projected ones as red bar histograms.  }
\label{pproj08mo}
\end{figure}
 
We ran a similar analysis for stars in the mass bin
0.6-0.9~M$_\odot$. The results are shown in Figure~\ref{pproj08mo} and
the model parameters are listed in Table~\ref{mod}. Starting from
the 5~Myr period distribution, a short coupling timescale between the
core and the envelope ($\tau_c$=1~Myr) is required to yield fast rotators
on the ZAMS, while a much longer coupling timescale ($\tau_c$=50~Myr)
is needed to account for the slow rotators.  As for the solar-mass
stars, the best results are obtained by assuming a rotation-dependent
coupling timescale. Figure~\ref{pproj08mo} shows that a model assuming
$\tau_c=500 \cdot(\omega/\omega_\odot)^{-1}$~Myr provides a reasonable
description of the observed angular momentum evolution of
0.6-0.9~M$_\odot$ stars from 5 to 130~Myr. A small fraction of ZAMS
stars in this mass range, however, have rotational periods longer than
10 days and up to $\sim$20 days, that are not accounted for by the
model. Assuming these slow rotators result from an extended
disk-locking period during the PMS, this suggests that disk lifetimes
may exceed 5~Myr for about 7\% of stars in this mass range, a result
that would be consistent with current estimates of accretion disk
lifetimes (Fedele et al. 2010).

\begin{figure}
\centering
\includegraphics[width=0.5\textwidth ]{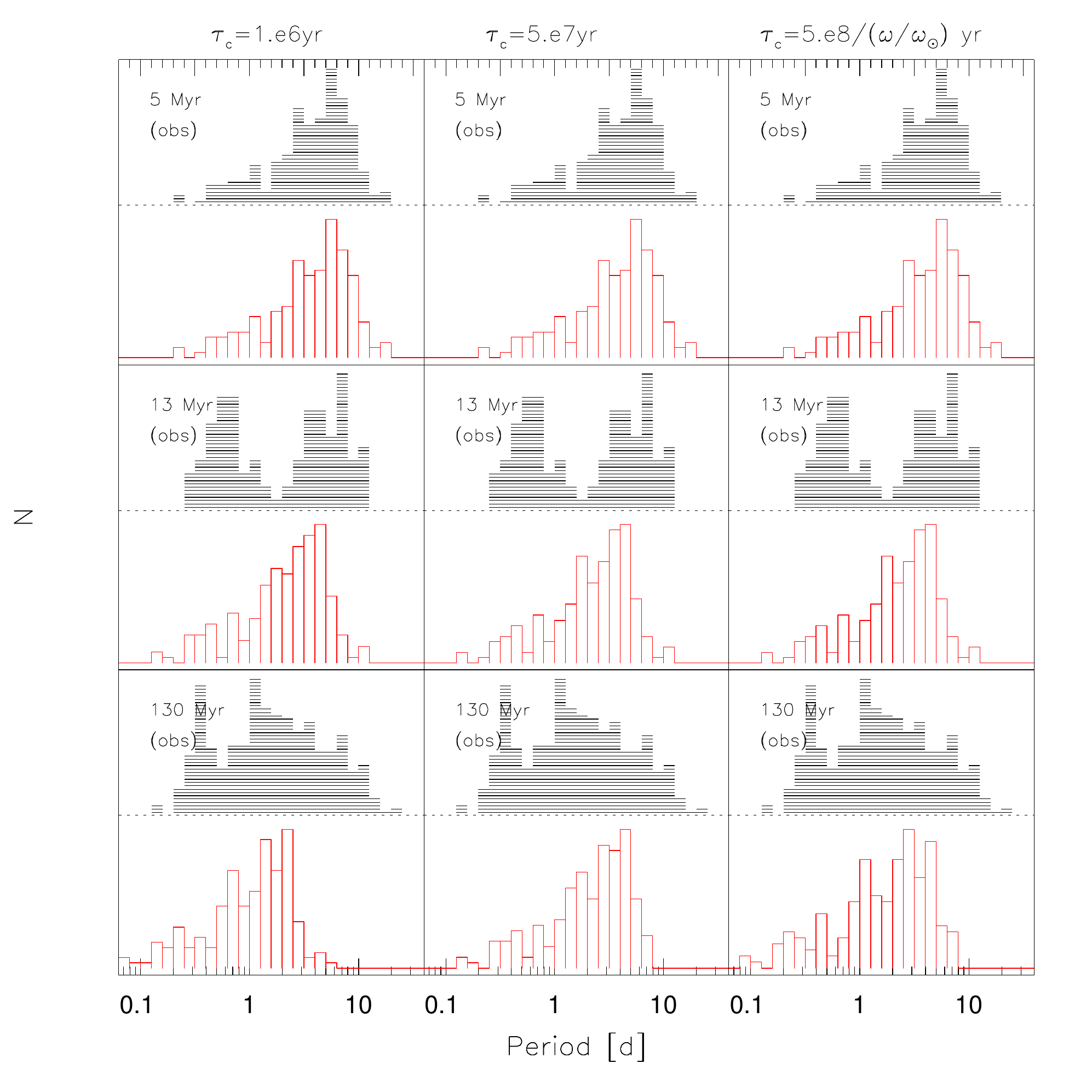}
\caption{ The initial distribution of periods at 5~Myr (top panels)
  for 0.4-0.6~M$_\odot$ stars is projected to 13 and 130 Myr (top to
  bottom panels) using 3 models : strong core-envelope coupling
  ($\tau_c=1$~Myr, left panels), weak core-envelope coupling
  ($\tau_c=50$~Myr, middle panels), and a velocity-dependent coupling
  timescale ($\tau_c=500 \cdot(\omega/\omega_\odot)^{-1}$~Myr, right
  panels). The observed distributions are shown as black shaded
  histograms, the projected ones as red bar histograms. }
\label{pproj05mo}
\end{figure}
 
Finally, model simulations were run for the lowest mass bin,
0.4-0.6~M$_\odot$. The results are shown in
Figure~\ref{pproj05mo}. While a short coupling timescale
($\tau_c$=1~Myr) clearly fails to reproduce slow rotators in this mass
range on the ZAMS, the observed evolution of the rotational period
distributions can be reproduced by assuming either a longer coupling
timescale ($\tau_c$=50~Myr) for all stars or a rotation-dependent
core-envelope coupling ($\tau_c=500
\cdot(\omega/\omega_\odot)^{-1}$~Myr). This is because, as the
convective envelope thickens in the lowest mass bin, surface rotation
becomes less sensitive to varying coupling timescales. As in the
0.6-0.9~M$_\odot$ bin, a small fraction (6\%) of low-mass stars have
rotational periods in excess of 10 days on the ZAMS, which may
indicate a prolonged star-disk coupling during the early PMS, up to
10~Myr. Alternatively, we cannot exclude that some of these slow
rotators, which mostly arise from the M50 cluster study by Irwin et
al. (2009), may be slowly-rotating field star contaminants.

Applying quantitative statistical analysis, such as 2-sided
Kolmogorov-Smirnov tests, indicates that the formal agreement between
the observed and model projected rotational distributions is
poor. This may result both from shortcomings of the models and from
uncertainties in the observed rotational distributions. As to the
models, we considered the accretion process to be fully completed by
5~Myr. As discussed above, assuming instead that a small fraction of
PMS stars have disk lifetimes in excess of 5~Myr would increase the
fraction of slow rotators on the ZAMS and bring the observed and
projected distribution in better agreement. As to the observed
distribution themselves, they may suffer various biases. We have shown
in Section~\ref{sec:bin} that short-period synchronized binaries
significantly impact on the h~Per rotational distribution and
consequently removed them from the analysis. To our knowledge,
synchronized binaries have not been identified and similarly discarded
in other studies. Field contaminants in the photometric samples may
also affect the rotational distributions by increasing the fraction of
slow rotators, with long period field stars being uncorrectly
classified as cluster members. Finally, environmental conditions may
play a role in shaping the initial distribution of angular momenta
(e.g. Clarke \& Bouvier 2000; Littlefair et al. 2010). Such intrinsic
cluster-to-cluster differences would undermine a detailed comparison
between observed and projected distributions. Given these numerous
sources of uncertainty, we believe it is premature to attempt to
reproduce the detailed shape of the observed rotational
distributions. Instead, we merely aimed here at a qualitative agreement
between the observed and modeled distributions by trying to best
account for the {\it range} of observed rotational periods as a
function of age.

\section {Conclusions}

We have presented the results of an extensive photometric monitoring
campaign in the 13~Myr open cluster h~Per to look for variability in
the $I$-band. Thanks to the combination of CFHT/MegaCam and Maidanak
data, the study is sensitive to periodic variations on timescales of
less than 0.2 day to 20 days. Selection of candidate members using
empirical isochrones in various CMDs in the optical and the
near-infrared found 2287 candidates in the Maidanak camera field of
view (18'$\times$18'). The field contamination is estimated to be
around 18\% according to the Besan\c{c}on Galactic model.

The light curves of the candidate cluster members were searched for
periodic modulations due to stellar rotation, giving 586 detections
over the mass range $0.4\le M/M_{\odot} \le 1.4$. This provides a
statistically large sample with uniform completeness of rotational
periods for low mass stars at 13 Myr, a time step that was not covered
by previous cluster studies. This age is particularly important to
understand the angular momentum evolution prior to the ZAMS since it
corresponds to the time when most of the stars have dissipated their
disk and start to freely spin-up as they contract towards the ZAMS.

The rotation period distribution exhibits a wide dispersion, with most
of the measured periods in the range $\sim 0.3$ to $\sim 9$ days and a
few slow rotators around 15 days. We found that photometric binaries
skew the distribution towards short periods, especially in the mass
range above 1.2 $M_{\odot}$. We suggest that this is due to the
presence of synchronized binaries that have a specific rotational
evolution due to tidal fields and should therefore be removed to
investigate the angular momentum evolution of single stars. The
distribution of the remaining 508 rotational periods (excluding the
photometric binaries with a period P$<1$d) does not show any clear
dependence with mass. In particular, the upper and lower envelopes
(corresponding respectively to slow and fast rotators) is remarkably
flat over the whole mass range.

Comparing the rotational period distribution of h~Per members to those
of low mass stars in other young clusters allowed us to model the
angular momentum evolution of low-mass stars during the PMS up to the
ZAMS. Models suggest that core-envelope decoupling occurs on a
timescale inversely proportional to surface rotation. Furthermore,
models indicate that less than 10\% of stars may remain coupled to
their disk beyond 5~Myr in order to reproduce the spin evolution to
the ZAMS, in agreement with current estimates of disk lifetimes.

\begin{acknowledgements}
  The authors thank B. Reipurth for giving us access to his CFHT data;
  K. Zwintz for the pulsation analysis of the 2 fast rotators;
  A. Scholz for useful discussion about fast rotation from
  synchronized binaries; F. Mignard for help with the FAMOUS program;
  and A. Robin for the use of the Besancon Galactic Model. We are also
  grateful to the CFHT QSO team for service observing and to
  J.-C. Cuillandre for the pre-reduction of CFHT data. This work has
  been supported by the EGIDE ECONET 1886YF program as well as by
  bilateral programs with the Academy of Science in Armenia and
  Ukraine. J. Bouvier and E. Moraux acknowledge funding from the
  Agence National Recherche grants ANR-2011-Blanc-SIMI5-6-020-01
  ``Toupies: Towards understanding the spin evolution of stars''
  ({http://ipag.osug.fr/Anr\_Toupies/}) and ANR-2010-JCJC-0501-1
  ``DESC: Dynamical Evolution of Stellar Clusters''
  ({http://ipag.osug.fr/~emoraux/DESC}) respectively.
\end{acknowledgements}


\begin{appendix} 

\section{Phased light curves of the 586 periodic h Per candidate members} 
 
\begin{figure*}
\includegraphics[width=1.0\textwidth]{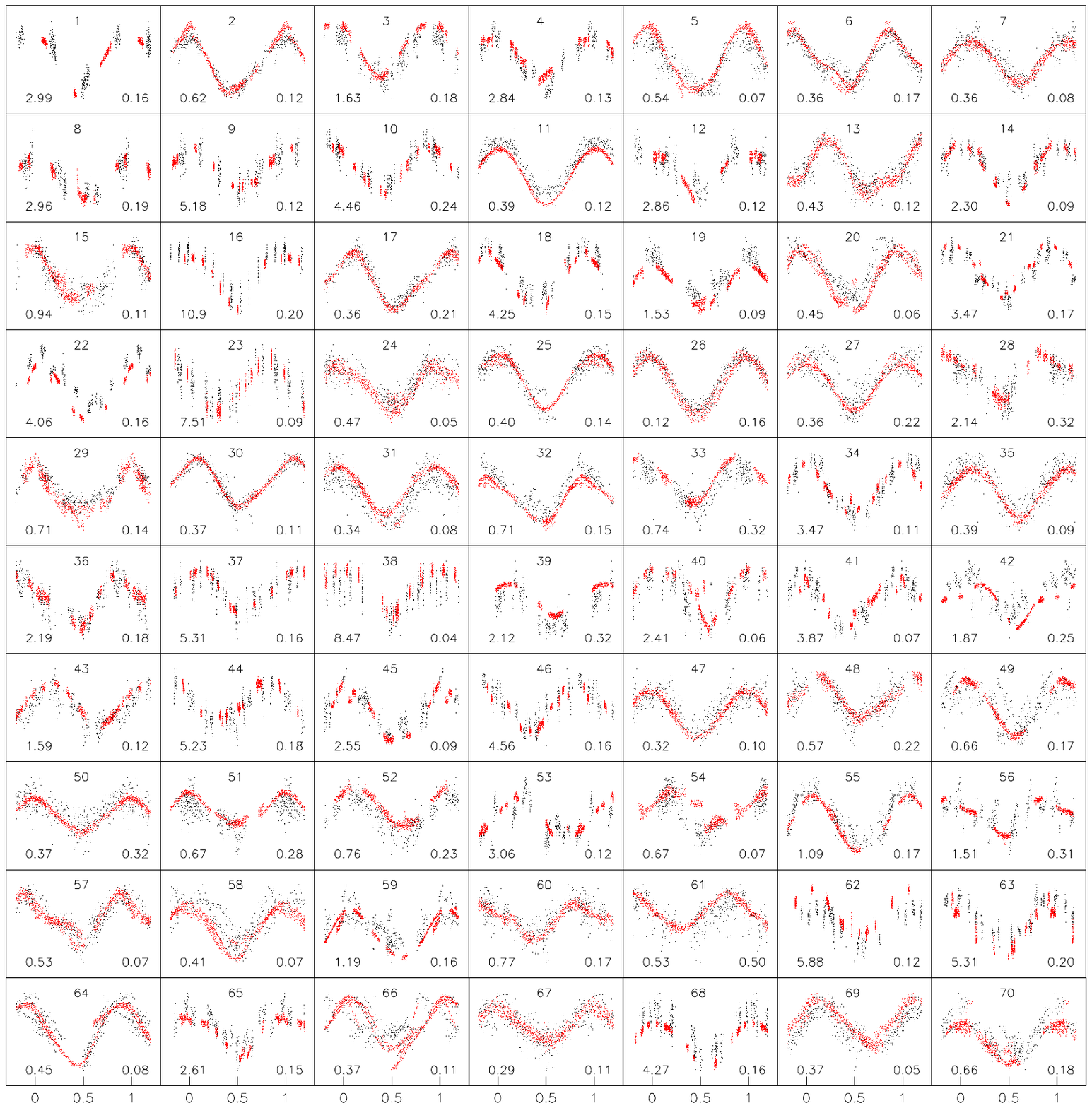}
\caption{The phased light curves of periodic h~Per candidate members
  (red dots : CFHT, back dots : Maidanak). The object number is given
  in each panel (top) as well as the period (in days, bottom left)
  and amplitude (in magnitude, bottom right). The light-curves are ordered by decreasing
  periodogram peak power.}
  \label{atlas1}
\end{figure*}
\begin{figure*}
\includegraphics[width=1.0\textwidth]{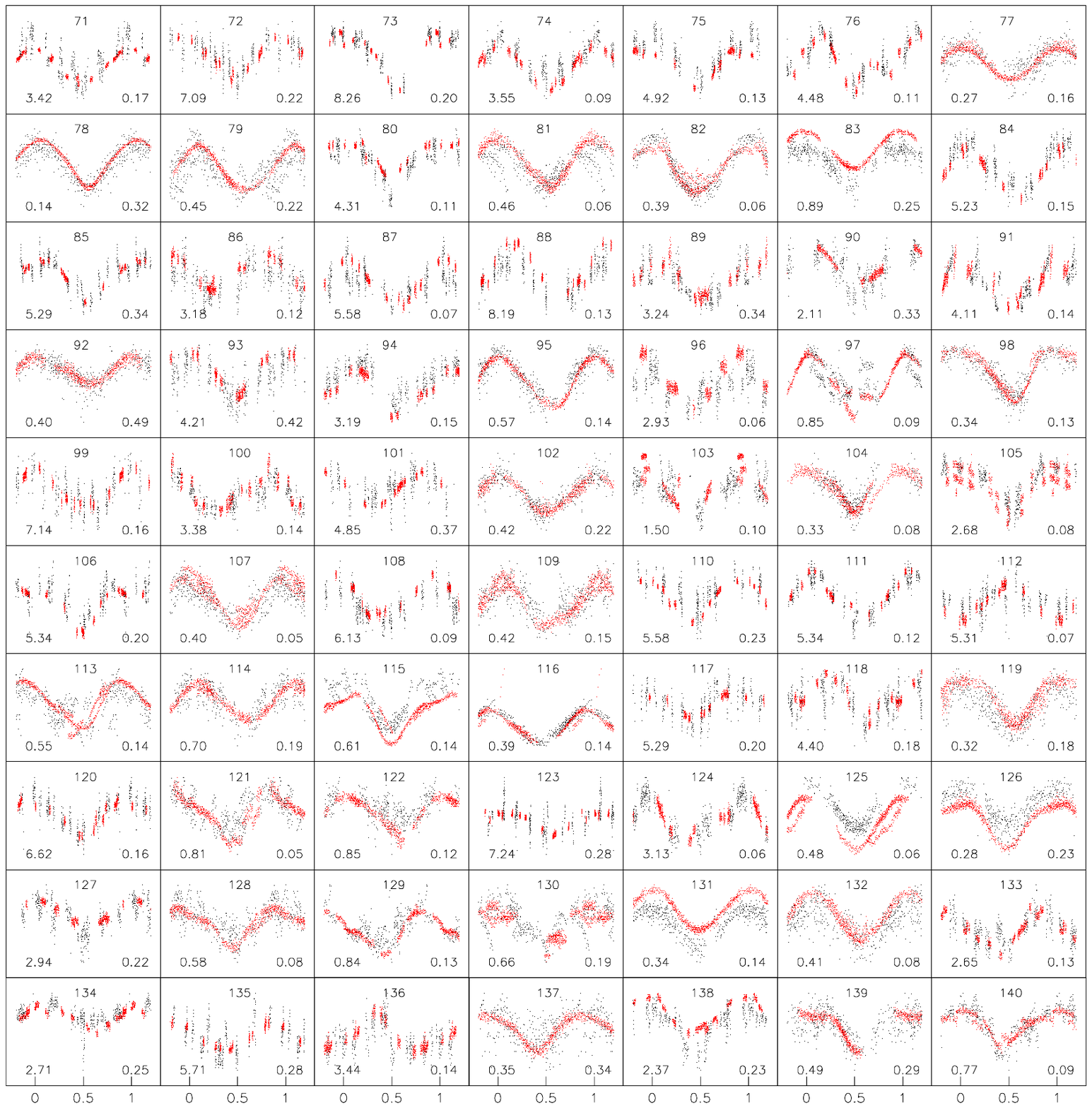}
 \caption{Fig.~\ref{atlas1} ct'd.}
 \label{atlas2}
\end{figure*}
\begin{figure*}
\includegraphics[width=1.0\textwidth]{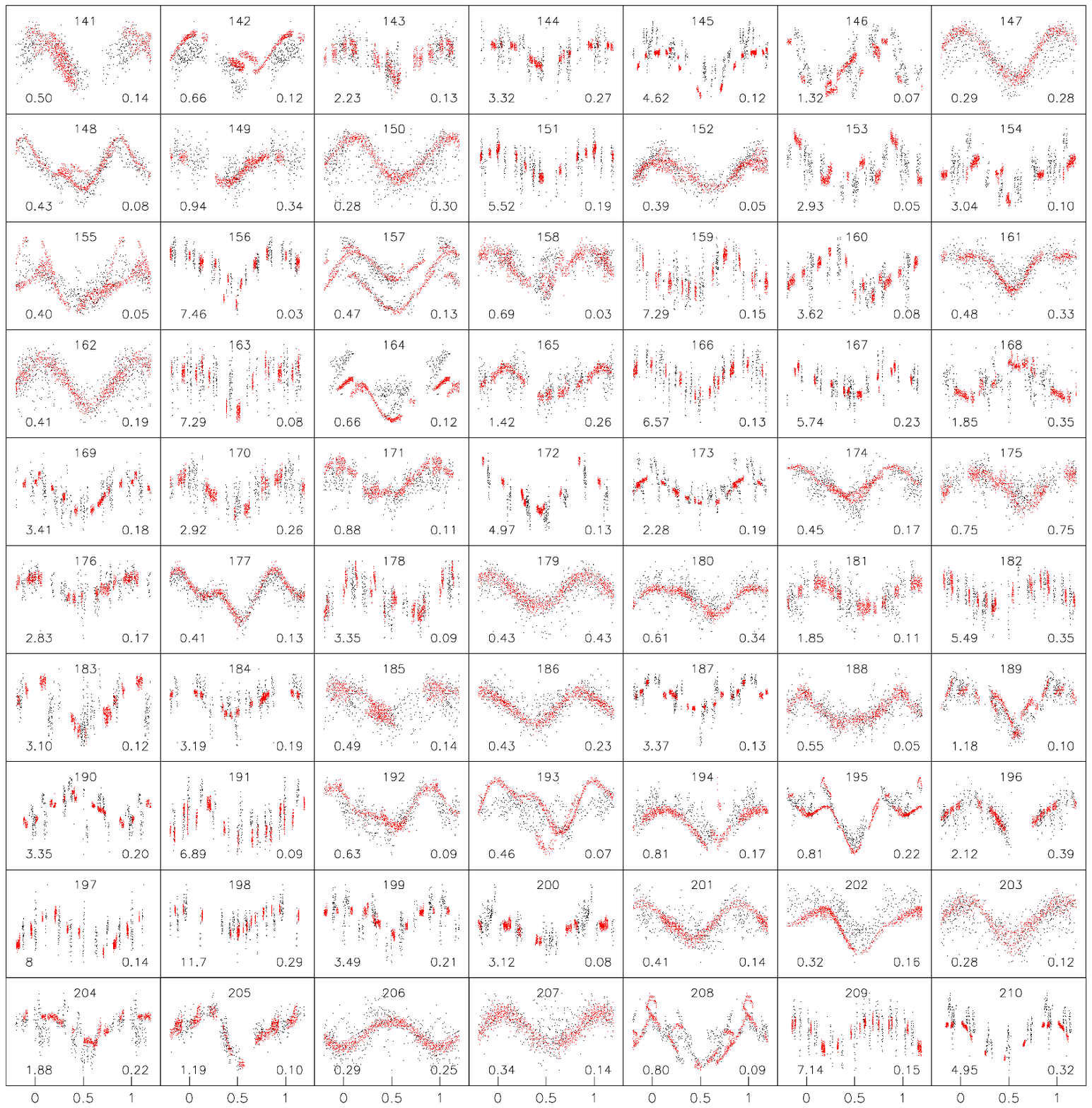}
 \caption{Fig.~\ref{atlas1} ct'd.}
 \label{atlas3}
\end{figure*}
\begin{figure*}
\includegraphics[width=1.0\textwidth]{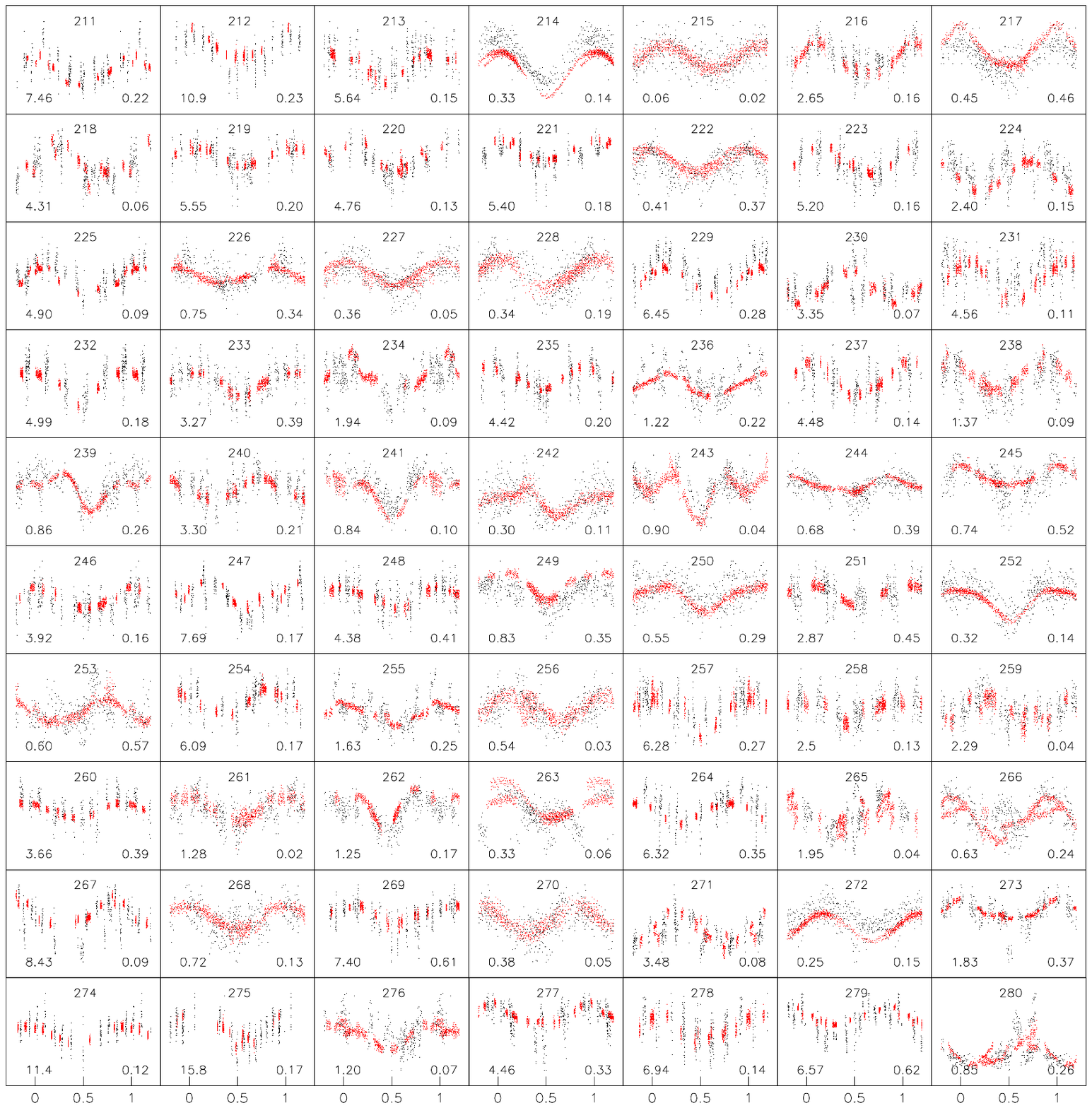}
 \caption{Fig.~\ref{atlas1} ct'd.}
 \label{atlas4}
\end{figure*}
\begin{figure*}
\includegraphics[width=1.0\textwidth]{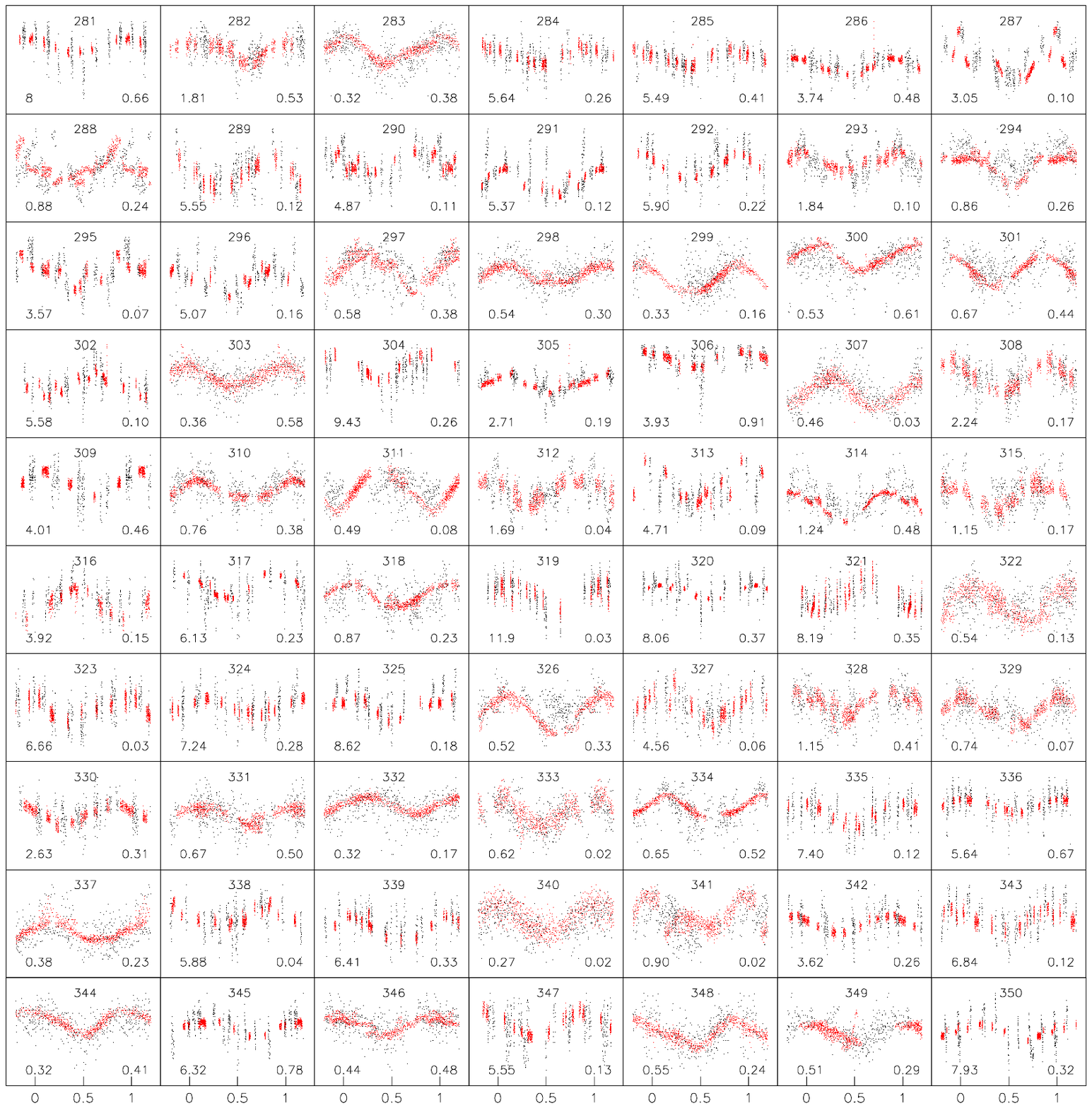}
 \caption{Fig.~\ref{atlas1} ct'd.}
 \label{atlas5}
\end{figure*}
\begin{figure*}
\includegraphics[width=1.0\textwidth]{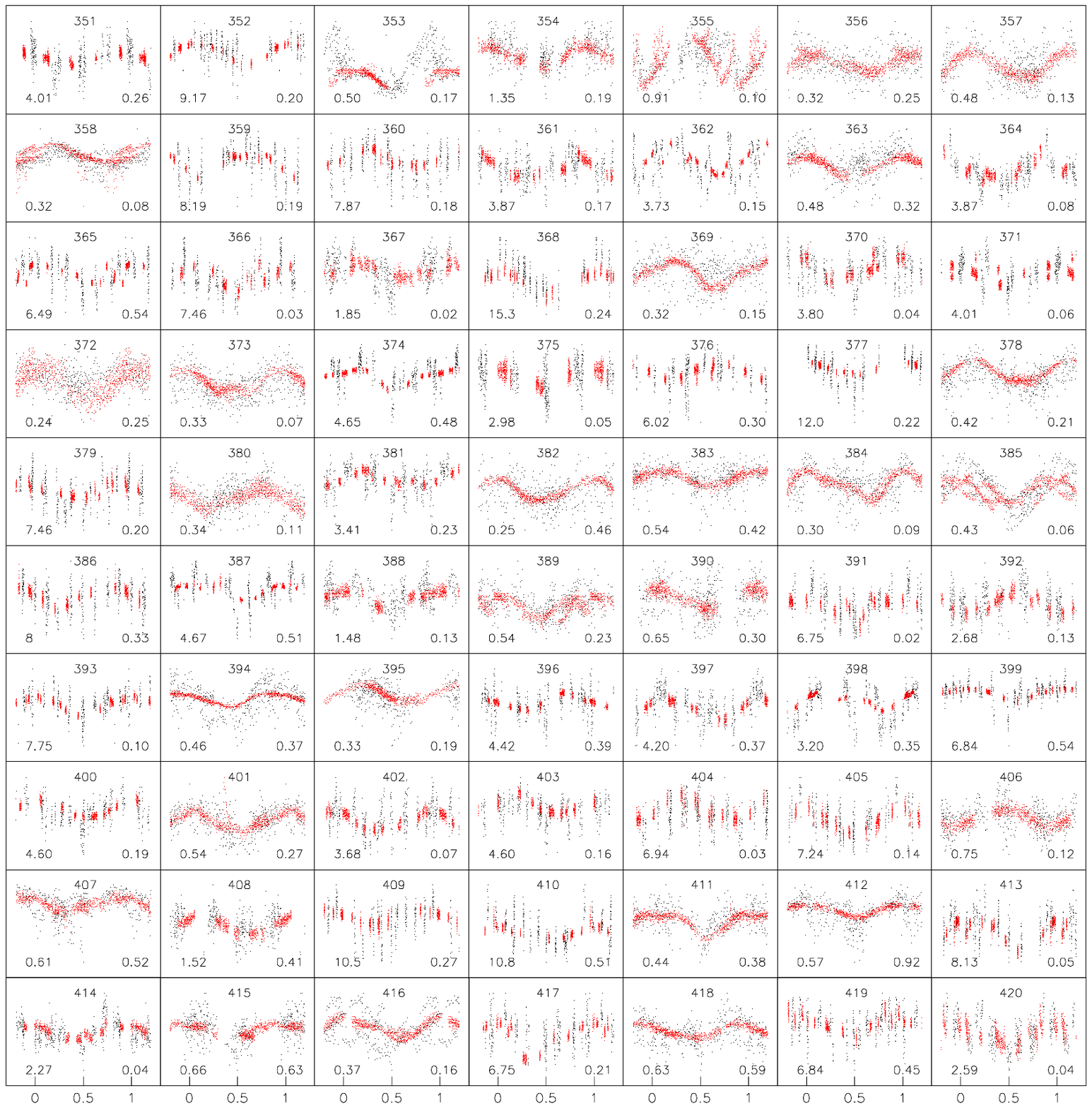}
 \caption{Fig.~\ref{atlas1} ct'd.}
 \label{atlas6}
\end{figure*}
\begin{figure*}
\includegraphics[width=1.0\textwidth]{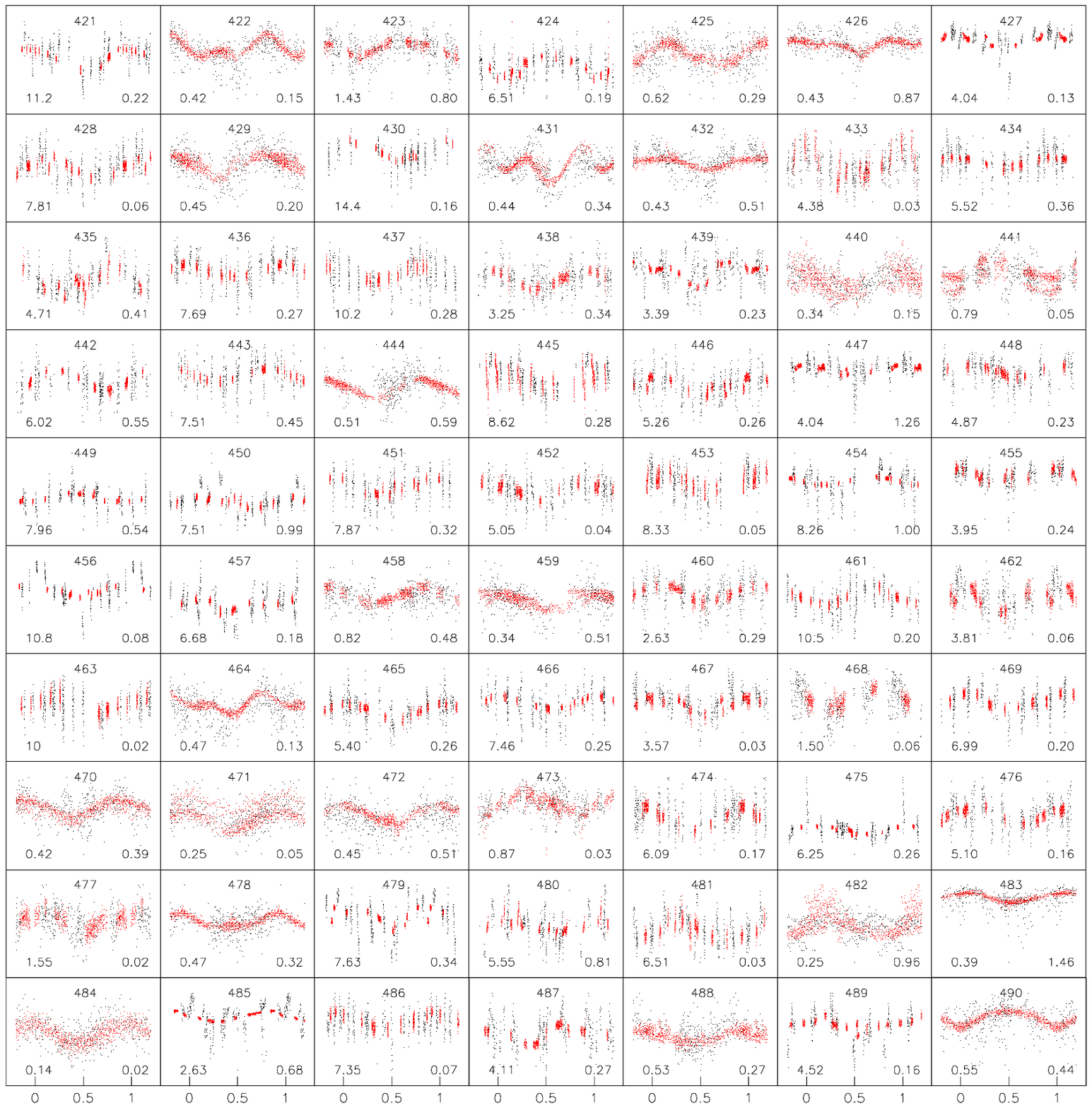}
 \caption{Fig.~\ref{atlas1} ct'd.}
 \label{atlas7}
\end{figure*}
\begin{figure*}
\includegraphics[width=1.0\textwidth]{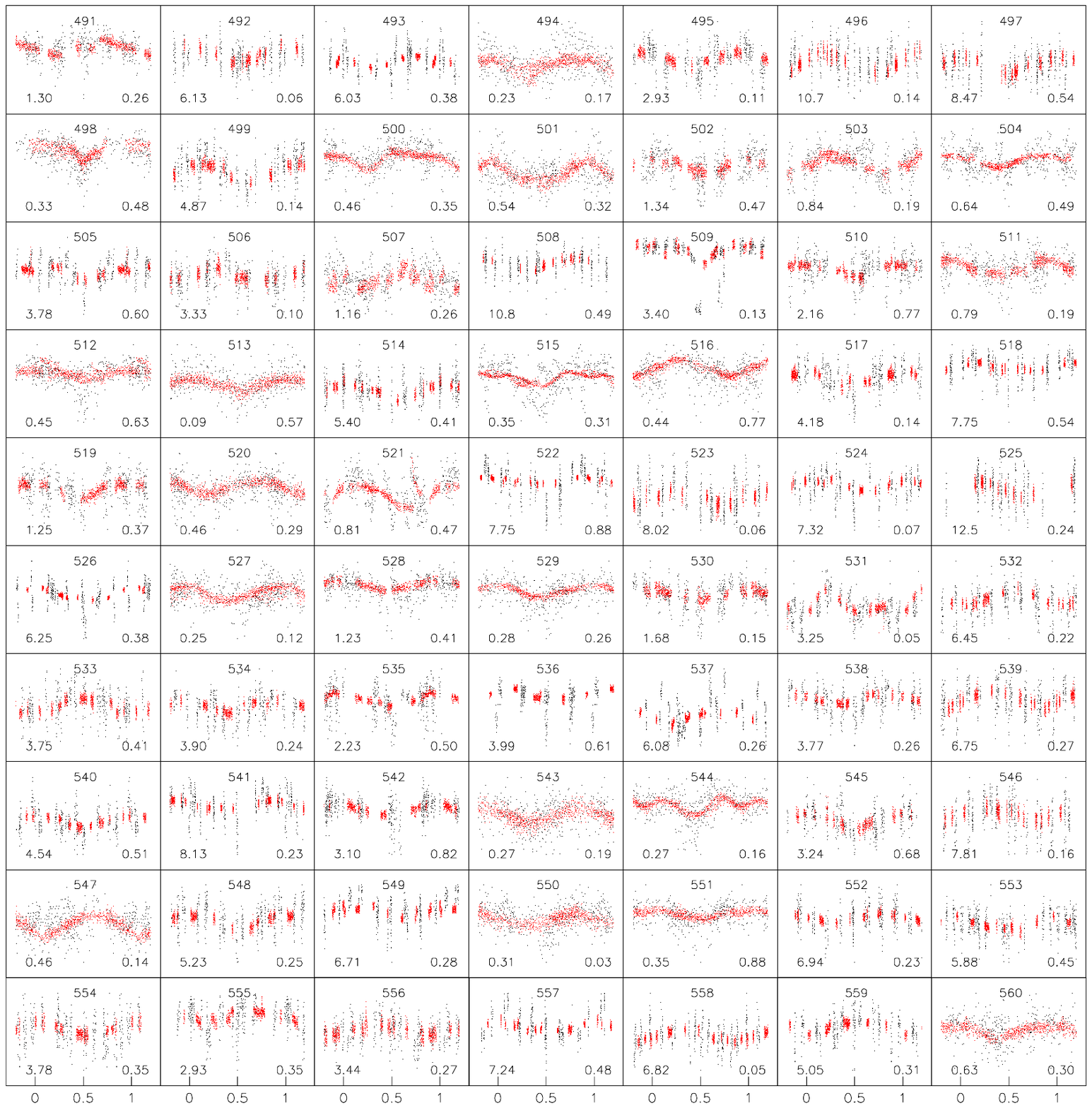}
 \caption{Fig.~\ref{atlas1} ct'd.}
 \label{atlas8}
\end{figure*}
\begin{figure*}
\includegraphics[width=1.0\textwidth]{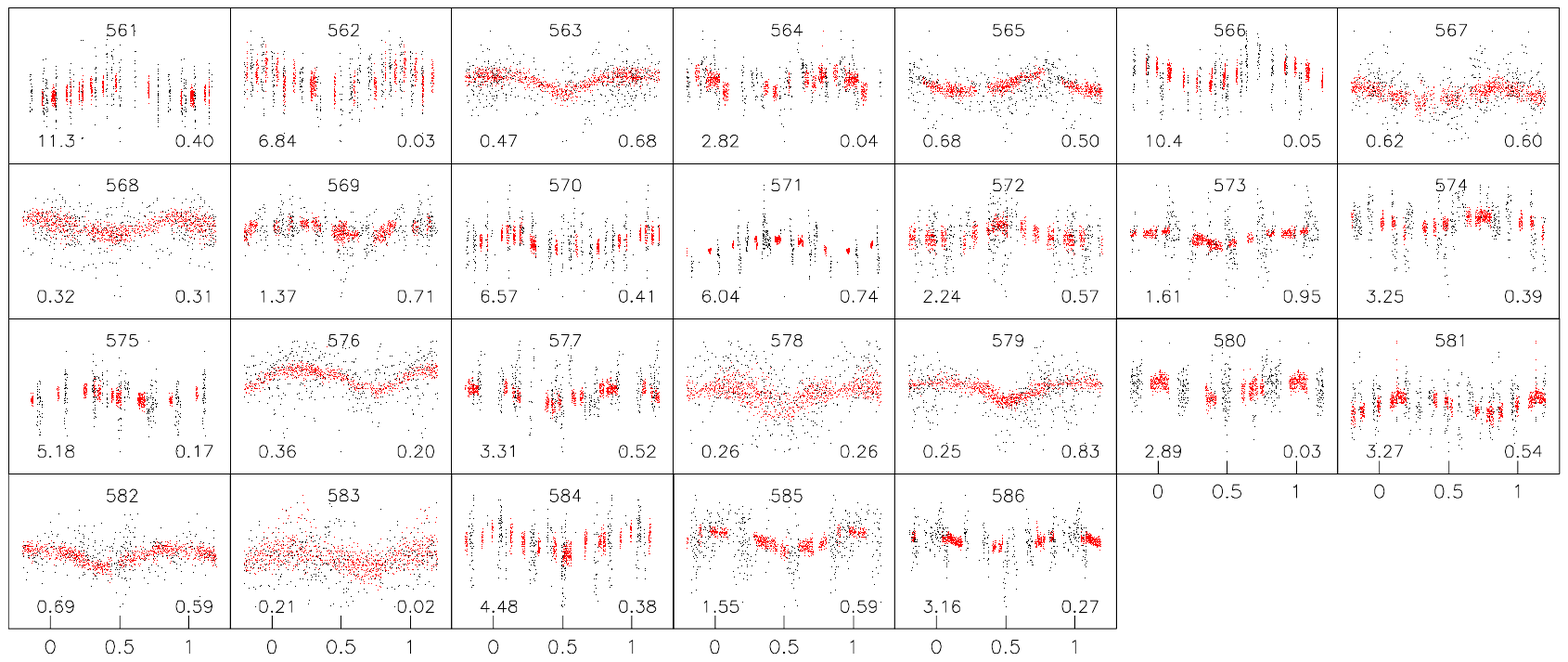}
 \caption{Fig.~\ref{atlas1} ct'd.}
 \label{atlas9}
\end{figure*}

\end{appendix}

\end{document}